\pdfoutput=1
\documentclass[11pt]{iopart}

\usepackage{amsfonts}
\usepackage{amssymb}
\usepackage{graphicx}
\usepackage{svg}
\usepackage{epstopdf}
\usepackage{verbatim}
\usepackage{color}
\usepackage{multirow}
\usepackage{mathcomp}
\usepackage{array}
\usepackage{bm}
\usepackage{wasysym}
\usepackage{setstack}
\usepackage{comment}
\usepackage[utf8]{inputenc}
\usepackage[T1]{fontenc}
\usepackage{textgreek}
\usepackage{csquotes}
\usepackage{harvard}
\usepackage{lineno}
\usepackage[normalem]{ulem}
\usepackage{float}
\usepackage{wrapfig}
\usepackage{xspace}
\usepackage[labelfont=bf,font=small,skip=3pt,margin=12mm]{caption}
\usepackage[thickspace,amssymb]{SIunits}

\newcommand{\betaplus}{$\beta^{\mathrm{+}}$}
\newcommand{\protheramon}{\texttt{ProTheRaMon}\xspace}

\usepackage{xcolor}
\usepackage{subfigure}
\usepackage[thickspace,amssymb]{SIunits}
\usepackage{todonotes}

\providecommand{\keywords}[1]
{
  \small	
  \textbf{\textit{Keywords---}} #1
}

\begin{document}
\textcolor{blue}{ Version date: \today }

\title[Feasibility of the J-PET to monitor range of therapeutic proton beams]{Feasibility of the J-PET to monitor range of therapeutic proton beams}

\author{Jakub Baran$^{1,2,3*}$, 
Damian Borys$^{4,5,6}$, 
Karol Brzeziński$^{6,7}$, 
Jan Gajewski$^{6}$, 
Michał Silarski$^{1,2,3}$, 
Neha Chug$^{1,2,3}$,
Aurélien Coussat$^{1,2,3}$,
Eryk Czerwi{\'n}ski$^{1,2,3}$,
Meysam~Dadgar$^{1,2,3}$,
Kamil~Dulski$^{1,2,3}$,
Kavya~V.Eliyan$^{1,2,3}$,
Aleksander Gajos$^{1,2,3}$,
Krzysztof~Kacprzak$^{1,2,3}$,
\L{}ukasz~Kap\l{}on$^{1,2,3}$,
Konrad~Klimaszewski$^{8}$,
Pawe\l{}~Konieczka$^{8}$,
Renata~Kopeć$^{6}$,
Grzegorz~Korcyl$^{1,2,3}$,
Tomasz~Kozik$^{1,2,3}$,
Wojciech~Krzemie{\'n}$^{9}$,
Deepak~Kumar$^{1,2,3}$,
Antony~J.Lomax$^{12,13}$,
Keegan~McNamara$^{12,13}$,
Szymon~Nied{\'z}wiecki$^{1,2,3}$,
Pawe\l{}~Olko$^{6}$,
Dominik~Panek$^{1,2,3}$,
Szymon~Parzych$^{1,2,3}$,
Elena~Perez~del~Rio$^{1,2,3}$,
Lech~Raczy{\'n}ski$^{8}$,
Moyo Simbarashe$^{1,2,3}$,
Sushil~Sharma$^{1,2,3}$,
Shivani$^{1,2,3}$,
Roman~Y.~Shopa$^{8}$,
Tomasz~Skóra$^{14}$,
Magdalena~Skurzok$^{1,2,3}$,
Paulina~Stasica$^{6}$,
Ewa~{\L{}}.~St\k{e}pie{\'n}$^{1,2,3}$,
Keyvan~Tayefi$^{1,2,3}$,
Faranak~Tayefi$^{1,2,3}$,
Damien~C.Weber$^{10,11,12}$,
Carla~Winterhalter$^{12,13}$,
Wojciech~Wi{\'s}licki$^{8}$,
Pawel Moskal$^{1,2,3}$, 
Antoni Rucinski$^{6}$} 
\address{$^1$Faculty of Physics, Astronomy and Applied Computer Science, Jagiellonian University, 30-348 Kraków, Poland}
\address{$^2$Total-Body Jagiellonian-PET Laboratory, Jagiellonian University, 30-348 Kraków, Poland}
\address{$^3$Center for Theranostics, Jagiellonian University, Kraków, Poland}
\address{$^4$Silesian University of Technology, Department of Systems Biology and Engineering, Gliwice, Poland}
\address{$^5$Biotechnology Centre, Silesian University of Technology, Gliwice, Poland}
\address{$^6$Institute of Nuclear Physics Polish Academy of Sciences, 31-342 Kraków, Poland}
\address{$^7$Instituto de F\'isica Corpuscular (IFIC), CSIC-UV, Valencia, Spain}
\address{$^8$Department of Complex Systems, National Centre for Nuclear Research, Otwock-Świerk, Poland}
\address{$^9$High Energy Physics Division, National Centre for Nuclear Research, Otwock-Świerk, Poland}
\address{$^{10}$Department of Radiation Oncology, Inselspital, Bern University Hospital, University of Bern, Bern, Switzerland}
\address{$^{11}$Department of Radiation Oncology, University Hospital of Zürich, Zürich, Switzerland}
\address{$^{12}$Centre for Proton Therapy, Paul Scherrer Institute, Villigen, Switzerland }
\address{$^{13}$Physics Department, ETH Z{\"u}rich, Z{\"u}rich, Switzerland}
\address{$^{14}$National Oncology Institute, National Research Institute, Krakow Branch, Krakow, Poland}

\eads{(corresponding author) *\mailto{jakubbaran92@gmail.com}}

\begin{abstract}

\textbf{Objective:}
The aim of this work is to investigate the feasibility of the Jagiellonian Positron Emission Tomography (J-PET) scanner for intra-treatment proton beam range monitoring.

\textbf{Approach:}
The Monte Carlo simulation studies with GATE and PET image reconstruction with CASToR were performed in order to compare six J-PET scanner geometries (three dual-heads and three cylindrical). We simulated proton irradiation of a PMMA phantom with a Single Pencil Beam (SPB) and Spread-Out Bragg Peak (SOBP) of various ranges. The sensitivity and precision of each scanner were calculated, and considering the setup's cost-effectiveness, we indicated potentially optimal geometries for the J-PET scanner prototype dedicated to the proton beam range assessment.

\textbf{Main results:}
The investigations indicate that the double-layer cylindrical and triple-layer double-head configurations are the most promising for clinical application. We found that the scanner sensitivity is of the order of 10$^{-5}$ coincidences per primary proton, while the precision of the range assessment for both SPB and SOBP irradiation plans was found below 1 mm. 
Among the scanners with the same number of detector modules, the best results are found for the triple-layer dual-head geometry. 

\textbf{Significance:} We performed simulation studies demonstrating that the feasibility of the J-PET detector for PET-based proton beam therapy range monitoring is possible with reasonable sensitivity and precision enabling its pre-clinical tests in the clinical proton therapy environment. Considering the sensitivity, precision and cost-effectiveness, the double-layer cylindrical and triple-layer dual-head J-PET geometry configurations seem promising for the future clinical application. Experimental tests are needed to confirm these findings.

\end{abstract}


\keywords{J-PET, PET, range monitoring, proton radiotherapy}

\section{Introduction}
\label{sec:Intro}


Radiation therapy is frequently applied as part of cancer treatment. Radiation therapy using accelerated protons offers an excellent depth dose distribution, characterized by maximum dose deposition at the end of the proton range~\cite{Paganetti2012}. This enables increasing the dose to the target volume, while reducing the dose to Organs-At-Risk (OAR), which, for selected indications, results in a reduction of late side effects. These dosimetric properties make proton radiation therapy the most popular choice of treatment~\cite{nystrom2020treatment} for deeply seated brain and head and neck tumors~\cite{durante2017charged}.

Apart from the advantages of proton radiotherapy, there are also some limitations. One of the major ones is the uncertainty in the proton beam range, which may lead to the underexposure of the target volume or overexposure of the OAR to the therapeutic dose~\cite{Paganetti2012}. One approach to overcoming this problem is the \textit{in-vivo} verification of the radiation delivery by monitoring beam range in real-time, providing information that may motivate the interruption of the beam delivery once the proton range differs from that in the prescribed treatment plan. Another option is to monitor the beam range after the treatment and to compensate for any deviations from the planned dose distribution by treatment adaptation~\cite{parodi2020latest,parodi2022experience}. As the primary protons are fully stopped within the patient's body, measurement of the secondary particles is used in these monitoring techniques. For these applications, prompt gamma imaging~\cite{krimmer2018prompt,pennazio2022proton}, secondary charged particles tracking~\cite{traini2019review,Battistoni2015,marafini2017mondo,bashkirov2017inbeam} and positron emission tomography (PET) techniques~\cite{Bauer2013,Bisogni2016,Ferrero2018,lang2022towards} have been proposed and tested pre-clinically and clinically.

The application of PET for proton range monitoring consists of imaging \betaplus-emitting isotopes, such as $^{11}C$ ($T_\frac{1}{2}$=20.4 minutes), $^{10}C$ ($T_\frac{1}{2}$=19 seconds) or $^{15}O$ ($T_\frac{1}{2}$=2 minutes), produced during the nuclear interactions of protons with the tissues in the patient. In general, four different PET acquisition protocols for range monitoring are considered~\cite{kraan2015range,shakirin2011implementation}. The in-room protocol enables the acquisition of the PET signal just after the irradiation, in the treatment room, where the patient is transported from the irradiation gantry to the standalone PET scanner~\cite{zhu2011monitoring,min2013clinical}. The off-line approach is similar, but differs in that the PET acquisition is performed in a different room, freeing the treatment room for the irradiation of the next patients \cite{parodi2007patient,knopf2008quantitative,knopf2011accuracy,hishikawa2002usefulness,handrack2017sensitivity}. Due to the fast decay of short-lived \betaplus-emitting radioisotopes and the effect of biological washout, this approach suffers from the lowest count statistics. The in-beam approach is performed during the patient irradiation and is characterized by high noise levels, as all secondary particles introduce noise to the registered coincidences \cite{fiedler2010effectiveness,iseki2004range,miyatake2010measurement}. Additionally, the inter-spill acquisition mode was investigated for the synchrotron facilities, where the pauses between spills are long enough to acquire PET signal \cite{fiorina2021detection,Ferrero2018}. 

Considering the various PET acquisition protocols, different PET scanner geometries have been introduced and tested, both experimentally or using Monte Carlo simulations. The commercially available PET scanners are the first choice option for the in-room and off-line approaches. It assures the best image quality and the highest efficiency of the systems with no need for any hardware or software development \cite{handrack2017sensitivity}. However, next to the radioactive decay and biological wash-out, the limitation (specifically for the in-room application) is the room size and occupation time. It is crucial to enable unrestricted rotation of both proton gantry and patient couch between the irradiation of the subsequent treatment fields. For the inter-spill and in-beam acquisition protocols, full-ring scanners are not an option due to geometrical constraints. There are several geometrical considerations for in-beam and inter-spill PET scanner: (i) the beam delivery system (free nozzle rotation during the irradiation, PET system components out of the way of the beam), (ii) patient couch (free movement between the irradiation fields) and (iii) additional medical equipment (e.g. for the anesthesia procedure) put additional requirements on the PET system configuration. To overcome these constraints, unconventional PET scanners have been designed and tested for in-beam/inter-spill proton beam range verification. These included dual-head scanners \cite{Ferrero2018,fiorina2020detection,baran2019studies,rucinski2019investigations} or more sophisticated configurations such as the axially shifted, single-ring OpenPET \cite{tashima2016development,yamaya2017openpet} or the axially slanted full-ring and dual-ring \cite{yoshida2017development,crespo2006detector} configurations.

To meet the requirements placed on PET-based range monitoring systems in proton radiation therapy, the Jagiellonian Positron Emission Tomography (J-PET) scanner~\cite{Moskal2021b,Moskal2021a,moskal2021simulating}, a novel, cost-effective, portable, modular PET scanner, based on plastic scintillator technology, is being considered for this application. 

Here, we present for the first time a feasibility study of the different J-PET geometries for the application of proton beam range verification. We performed Monte Carlo simulations in homogeneous media in order to compare between six geometries (three dual-heads and three cylindrical), which could be potentially considered for in-beam, inter-spill, in-room, and off-line beam range monitoring. We report the relative efficiency of the scanners for Single Pencil Beam (SPB) and Spread-Out Bragg Peak (SOBP) irradiation plans. Quantitative analysis is conducted to assess the precision of range detection in a uniform phantom of different scanner geometries and indicate the optimal J-PET configuration for proton beam range monitoring.

\section{Materials and Methods}
\label{sec:MM}

\subsection{J-PET scanner and geometries for proton therapy range monitoring}
\label{mm:geometries}

\begin{figure}
    \centering
    \begin{subfigure}{}
        \centering
        \includegraphics[width=.49\linewidth]{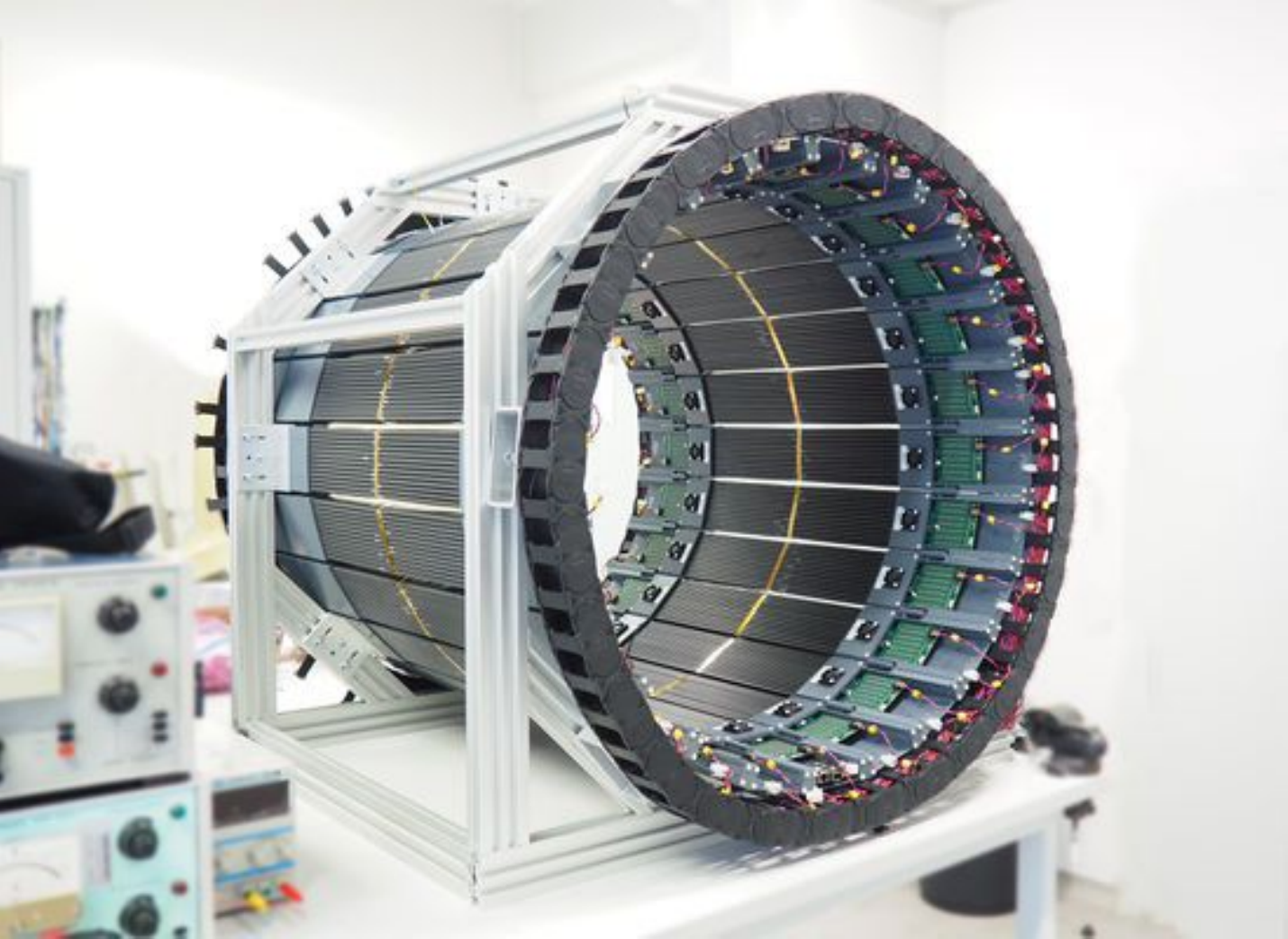}
        \includegraphics[width=.49\linewidth]{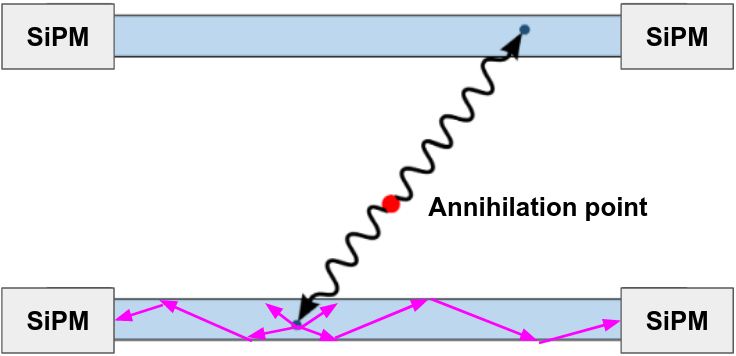}
    \end{subfigure}
    \caption{The modular J-PET scanner (left panel) developed as a cost-effective, diagnostic total-body PET prototype that was investigated in this study for proton therapy range monitoring. The presented geometry corresponds to the single layer cylindrical configuration (see Fig. 2A). The principle of annihilation gamma-rays detection with the J-PET module is illustrated in the right panel. Annihilation gamma-rays(black arrows) create in the plastic scintillator photons (magenta arrows) which propagate to the ends of the strip and are converted to the electronics signals by the silicon photomultipliers (SiPM).}
    \label{fig:modular_jpet_and_principle}
\end{figure}

The existing prototype of the J-PET scanner and the schematic illustration of its operation principle is presented in Figure~\ref{fig:modular_jpet_and_principle}. The annihilation gamma-rays leaving the imaged object to interact with the plastic scintillator strips. They deposit energy in the plastic mostly via Compton scattering. The scintillation light produced by these interactions propagates to the opposing ends of the scintillator strip where the light is converted into an electronic signal by silicon photomultipliers (SiPMs). The disadvantage of the plastic scintillators with respect to the conventionally used crystal scintillations is their efficiency~\cite{vandenberghe2020state} which in the case of the proton range monitoring is enhanced as the expected statistics is far beyond the level of clinical activities. Considering that the plastics scintillators are relatively cheap and have photomultipliers at their ends, the improvement in PET signal quality could be achieved by an increase of the thickness of the plastic scintillator, adding subsequent layers of plastic modules or increasing the length of the scintillators to enlarge the Field-Of-View (FOV)~\cite{moskal2016time}. Application of any of these approaches will increase the price of the system, however, the difference is not as substantial as for the organic scintillators~\cite{vandenberghe2020state,moskal2020prospects}. Taking into consideration, the relatively low price of the technology, its portability and the possibility to build various geometries with the same amount of modules, it makes J-PET scanner promising for proton range monitoring. Additionally, the J-PET could be also applied with success in Total-Body PET imaging~\cite{kowalski2018estimating,moskal2021simulating,moskal2020prospects}, multi-gamma tomography using e.g. positronium imaging~\cite{Moskal2021b,moskal2019feasibility,moskal2021positronium}, fundamental physic studies on quantum entanglement~\cite{hiesmayr2019witnessing,sharma2022decoherence}, studies of discrete symmetries in nature~\cite{Moskal2021a} or PET data reconstruction methods development~\cite{shopa2021optimisation,raczynski20203d}. Moreover, plastic scintillators in J-PET are characterized by short light signals with a decay constant of about 2 ns (factor of 20 to 150 less than these of crystal detectors)~\cite{moskal2021simulating}.  Therefore J-PET has by two orders of magnitudes lower chances for signal pile-ups  with respect to crystal PET systems, making it especially promising for the application of monitoring flash radiotherapy with high power radiation dose~\cite{vozenin2022towards}.  

The modular J-PET technology (as presented in Fig.~\ref{fig:modular_jpet_and_principle}) was developed to allow its reconfiguration for different applications. The J-PET module is built out of 13 separate 50\,cm long scintillator strips, each having a cross-section of 6$\times$24\,mm$^{2}$. Individual scintillators are covered with kapton and reflective foils~\cite{niedzwiecki2017j,kaplon2020technical}. Each of the 13 scintillation strips is connected to 8 SiPMs, 4 at each side of the strip, which convert the light into an electronic signal, further processed with the FPGA electronics~\cite{korcyl2018evaluation}.

\begin{figure}[h!]
\centering
{\includegraphics[width=.8\linewidth]{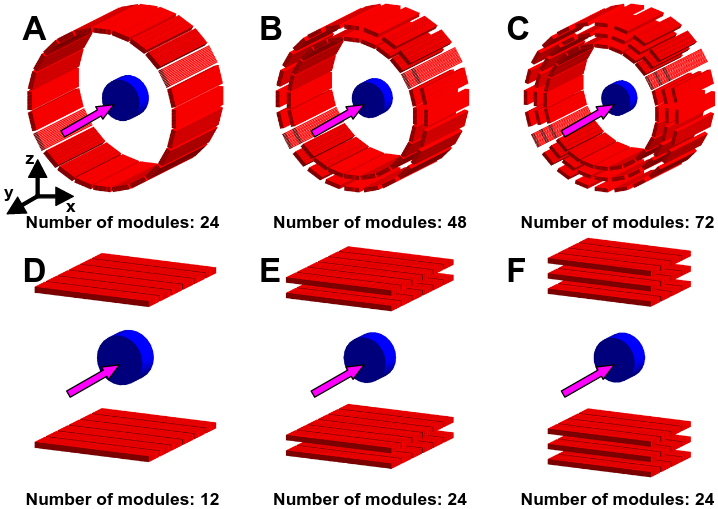}} 
\caption[]{The J-PET based geometrical configurations proposed for application in proton therapy range monitoring. In the Monte Carlo simulation study presented here we investigated: single layer cylinder (A), double layer cylinder (B), triple layer cylinder (C), single layer dual-head (D), double layer dual-head (E), and triple layer dual-head (F) configurations. A cylindrical water (blue) phantom was isocentrically positioned inside each of the configurations. The number of modules per layer in each head for single, double and triple layer dual-head setups is 6, 6 and 4, respectively. Purple arrows show the direction of the proton beam.}
\label{fig:gate_configurations_visualisation}
\end{figure}

For the purpose of proton therapy range monitoring, we propose and investigate six PET scanner configurations, built from the J-PET modules  (Figure~\ref{fig:gate_configurations_visualisation}). The PET geometries can be classified as two general types: the cylindrical (Figure~\ref{fig:gate_configurations_visualisation}A-C) and the dual-head (Figure~\ref{fig:gate_configurations_visualisation}D-F) configurations, each of them in a single-, dual-, and triple-layer geometry. 
The cylindrical setups could be used for the off-beam/in-room applications, whereas dual-head setups could be potentially also considered for the in-beam/inter-spill scenarios~\cite{shakirin2011implementation,parodi2020latest}. Each layer of the multi-layer cylindrical system consisted of 24 modules. The dual-head configurations consisted of 12, 24 and 24 modules for single, double and triple layer setups, respectively. The radius of the system, defined as the distance between the isocenter and the surface of the innermost strip in the module, was equal to 369.9\,mm and 300.0\,mm for cylindrical and dual-head configurations, respectively. The gap between adjacent layers was fixed to 44\,mm for all setups. 

The modules in the cylindrical configurations were positioned parallel to the beam direction. In contrast, the modules in dual-head configurations were positioned perpendicular to the beam, motivated by the potential improvement of the J-PET detector resolution in the direction perpendicular to the strips. When the modules are positioned parallel to the beam direction (cylindrical setups), the precision of the range measurement depends on the resolution of the interaction position along the strip, which is determined by the detector's timing properties~\cite{moskal2016time}. On the other hand, for the dual-head configurations, the range measurement depends on the resolution determined by the width of the plastic strips (6 mm), which is superior to that along the strip length. 



\subsection{Monte Carlo simulations}
\label{mm:simulation_setup}

In this work, we exploited Monte Carlo methods for simulation and evaluation of the feasibility of various J-PET geometries for proton therapy range monitoring. We used the \protheramon software framework, in which the delivery of proton therapy treatment plans were simulated, scoring the $\beta^+$ activity produced by protons during the treatment, as well as the J-PET scanner response to the annihilation gammas \cite{borys2022protheramon}. The \protheramon framework utilizes the GATE platform for Monte Carlo simulations~\cite{sarrut2022opengate}, the CASToR package for PET image reconstruction~\cite{merlin2018castor}, and in-house developed Python and bash scripts. The framework consists of five separate simulation and data processing steps.
\protheramon offers complete processing of proton therapy treatment plans, patient geometries based on Computed Tomography (CT), the simulation of treatment delivery and intra-treatment PET imaging, taking into account therapy and imaging coordinate systems, beam models specific to a given proton therapy facility, as well as activity decay during the treatment and PET imaging protocol \cite{borys2022protheramon}.


For simulations of the treatment plan irradiation, we used the Geant4 QGSP\_BIC\_HP\_EMY physics list with the RadioactiveDecay model. The beam model of the Cyclotron Centre Bronowice (CCB) Krakow proton therapy centre, along with the CT calibration, were implemented as described in \cite{Gajewski2021a}. For the simulation of the two 511 keV photon propagation from the positron-electron annihilation position to the PET scanner and their interaction with the plastic scintillator, the emlivermore\_polar physics list was used. We did not simulate scintillation processes, but we considered the energetic resolution of the scintillators \cite{moskal2021simulating} calculated based on experimental measurements, performed for the plastic scintillator strips~\cite{moskal2014test}. A 200-380\,keV energy window was set to extract a list of the coincidences, along with a 3\,ns time window \cite{moskal2021simulating}.

In this study, we simulated the irradiation of a PMMA phantom with proton SPB and SOBP of various ranges, to investigate the feasibility of different J-PET geometries for proton range monitoring. All treatment plans were optimized in the Varian Eclipse 16.1 treatment planning system, used routinely in CCB for patient treatment planning.

The SPB plans were prepared with nominal proton energies of 125.68\,MeV, 127.15\,MeV, 129.34\,MeV, and 132.25\,MeV, irradiating a $5\!\times\!20\!\times\!5$\,cm$^3$ PMMA phantom, achieving the corresponding Bragg peak range of 100\,mm, 102\,mm, 105\,mm, and 109\,mm, respectively. The plans were normalized to 8\,Gy 
in the Bragg peak maximum, which corresponds to about $3.3\cdot10^9$ primary protons.

Three sets of SOBP plans were prepared to obtain homogeneous dose distributions in $3\!\times\!3\!\times\!3$\,mm$^3$, $5\!\times\!5\!\times\!5$\,mm$^3$, and $7\!\times\!7\!\times\!7$\,mm$^3$ cubes, inside a $20\!\times\!15\!\times\!20$\,cm$^3$ PMMA phantom. Each of these sets was optimized to obtain the SOBP range the same as for the SPB plans, i.e.~100\,mm, 102\,mm, 105\,mm, and 109\,mm, leading to 12 SOBP plans of varying volume and range. The plans were normalized to 4\,Gy(RBE) in the SOBP cube, leading the total number of primaries of about $4.0\cdot10^{10}$, $9.6\cdot10^{10}$, and $2.1\cdot10^{11}$, for the small, medium, and large cubes, respectively. Both, the SPB and SOBP plans were irradiated along the y direction of the corresponding phantom (see Figure~\ref{fig:gate_configurations_visualisation}).

For the simulation of 511 keV annihilation photons propagation, the phantoms were positioned isocentrically inside the PET scanners. We assumed the in-room PET acquisition scenario, with a 90-second time delay from the end of the irradiation to the beginning of the PET acquisition, which lasted for 120 seconds. The available computing power enables performing simulations in a relatively short time. In total, the mean simulation time of the proton irradiation plans for SPB and consecutive SOBP plans from small to large cubes was 12 minutes, 7.2 hours, 16.8 hours, and 14.3 hours, respectively. Corresponding numbers for the PET acquisition simulation was 2.4, 31.5, 70.1, and 135.0 minutes.

\subsection{PET image reconstruction}
\label{mm:pet_recon}

The CASToR software~\cite{merlin2018castor} v. 3.1 was used for the PET data reconstruction. Since CASToR does not allow TOF modelling of the scintillation quanta propagation along the plastic scintillator strips, these were discretized to 100 artificial scintillators ($6\!\times\!24\!\times\!5$\,mm$^3$) along the longest dimension of the plastic strip. This discretization corresponds to the TOF resolution ($FWHM=5~mm$) along the the J-PET strips~\cite{smyrski2017measurement,moskal2021simulating}. The list-mode TOF MLEM reconstruction algorithm was used with TOF resolution equal to 500\,ps (FWHM) and with Siddon projector. Reconstructed images were corrected for sensitivity and attenuation. The PMMA linear attenuation coefficient was set to 0.104\,cm$^{-1}$. Reconstruction voxel was 2.5\,mm$^3$, isotropic. The reconstructed images were smoothed with the 3D Gaussian filter with $\sigma$=1\,voxel \cite{baran2019studies}. The images were reconstructed with 3 iterations.

\subsection{Analysis}
\label{mm:analysis}

The comparison of J-PET setup configurations was conducted considering (i) the sensitivity, and (ii) the precision of range shift detection. 

We defined the sensitivity, denoted $\eta$, as the ratio of the number of detected coincidence events per primary proton. It was calculated for each geometry and SPB/SOBP of different range. The $\eta$ factor is given as:

\begin{equation}
    \eta=\frac{c}{p},
\end{equation}

\noindent where $c$ is the number of registered coincidences and $p$ is the number of simulated primary protons.
In order to compare the six J-PET configurations, we defined and calculated the geometry-dependent normalized sensitivity factor $H$ given as:

\begin{equation}
    H=\frac{\eta_{geom}}{\eta_{ref}} \quad ,
\end{equation}

\noindent where the $\eta_{geom}$ is calculated for the investigated simulation setup (considering both SPB and SOBP simulations) and $\eta_{ref}$ is calculated for the single layer cylindrical geometry.

The mean value $\overline{H}$ was calculated separately for the SPB and SOBP studies for each geometry. For the SPB $\overline{H}$ was averaged over 4 beam ranges and for the SOBP over 4 beam ranges and 3 different dose cube sizes.




We also performed a quantitative analysis of the precision which can be expected for detecting proton beam range with different J-PET geometry configurations. 
The dose range, $R_D$, was calculated as the depth of 80\% of the distal fall-off of the integral depth dose (IDD) profile of SPB or central axis profile of SOBP.
Combinations of four dose ranges of SPB or SOBP, i.e.~$R_D$ were equal to 100\,mm, 102\,mm, 104\,mm, and 109\,mm, allowed to analyze six dose range differences, $\delta R_D$, i.e. 2\,mm, 3\,mm, 4\,mm, 5\,mm, 7\,mm, and 9\,mm. The differences in dose range, $\delta R_D$, of SPB or SOBP irradiations were assumed to be the reference for the activity range analysis. 

The activity range, $R_A$~\cite{camarlinghi2014beam}, was calculated by fitting a sigmoid function to the distal fall-off of the 3D activity distribution reconstructed with a given J-PET geometry. Based on the fit, the $R_A$ was determined at 50\% of the distal fall-off maximum~\cite{kraan2015first,kraan2015range,min2013clinical}. 
The difference of six activity ranges, $\delta R_A$, associated with the six reference differences in dose range, $\delta R_D$, was calculated by subtracting the range of fitted activity profiles.
The deviation between activity and dose range difference, $\Delta R$, was individually calculated for each geometry setup and for SPB and SOBP, as:
\begin{equation}
    \Delta R = \delta R_D - \delta R_A \quad .
\end{equation}
Note that for a PET scanner that flawlessly reconstructs the emission activity distribution, the $\Delta R$ is expected to be equal to zero. 

We calculated the uncertainty of the deviation between activity and dose range difference ($\Delta R$) for each investigated scanner geometry and for SPBs and SOBPs of different field size. For this purpose, the mean $\overline{\Delta R}$ and standard deviation $\sigma{\Delta R}$ of $\Delta R$ was calculated, $\sigma{\Delta R}$ being proposed as a metric of the precision of the activity range detection.


\section{Results}
\label{sec:Results}

\subsection{Sensitivity}

The ratio of detected coincidence events per primary proton ($\eta$), and the normalized sensitivity factor ($\overline{H}$), all computed for SPB and SOBP irradiations in six J-PET geometries are given in Tab.~\ref{tab:efficiences}. The $\eta$ factor for the SPB ranges from $0.4 \cdot 10^{-5}$ to $4.8 \cdot 10^{-5}$ for single layer dual-head and triple layer cylindrical setups, respectively. For the SOBP the numbers are smaller due to the bigger phantom (greater attenuation) and range from $0.25 \cdot 10^{-5}$ to $2.0 \cdot 10^{-5}$ for single layer dual-head and triple layer cylindrical setups, respectively. 
The greatest value of $\overline{H}$ is observed for the triple and double layer cylindrical setups and the lowest for the single layer dual-head. For the SBP irradiation, the sensitivity of the J-PET geometries consisting of the same number of modules, i.e., single layer cylindrical, double layer dual-head and triple layer dual-head is comparable within about 10\%. However, for the SOBP scenario, the geometry-specific sensitivity factor $\overline{H}$ varies significantly  (about 2-2.4 times) for double and triple layer dual-head setups with respect to the single layer cylindrical geometry. It shows the advantage of adding the subsequent detector layers over the greater coverage of the FOV when the same number of modules is available. For the cylindrical setups, the addition of the new layer of modules increases the sensitivity of the system.

\begin{table}[ht]
\caption{The sensitivity $\eta$ and normalized sensitivity factor $\overline{H}$ for SPB and SOBP irradiations computed for the investigated J-PET geometries. The reference geometry is the single layer cylindrical setup.}
\centering
\begin{tabular}{ |c|c|c|c|c|c|c| } 
\hline
 & \multicolumn{3}{c|}{SPB} & \multicolumn{3}{c|}{SOBP} \\ 
\cline{2-7}
Setup & $\eta [10^{-6}]$ & $\sigma(\eta) [10^{-6}]$ & $\overline{H}$ & $\eta [10^{-6}]$ & $\sigma(\eta) [10^{-6}]$ & $\overline{H}$ \\
\hline
Single layer cylindrical & 9.45 & 0.29 & 1.0 & 3.64 & 0.22 & 1.0 \\ 
Double layer cylindrical & 27.41 & 0.80 & 2.9 & 10.76 & 0.65 & 2.9 \\ 
Triple layer cylindrical & 45.72 & 1.26 & 4.8 & 18.00 & 1.11 & 5.0 \\ 
Single layer dual-head & 3.79 & 0.13 & 0.4 & 2.45 & 0.19 & 0.7 \\ 
Double layer dual-head & 10.55 & 0.35 & 1.1 & 7.21 & 0.56 & 2.0 \\ 
Triple layer dual-head & 10.22 & 0.26 & 1.1 & 8.92 & 0.78 & 2.4 \\
 \hline
\end{tabular}
\label{tab:efficiences} 
\end{table}



\subsection{Examples of reconstructed activity distributions and profiles}

We selected exemplary images for cylindrical and dual-head configurations considering the sensitivity ($\overline{H}$ factor) and cost-effectiveness of the setup. For the cylindrical setup, the $\overline{H}$ factor increases by about 300\% between single (24 modules) and double layer (48 modules) configuration, while increasing only by about 70\% from the double (48 modules) and triple layer (72 modules) configuration. We, therefore, consider the double layer geometry as the more cost-effective solution for the cylindrical configuration~\cite{borys2022protheramon,moskal2021simulating}. The triple layer dual-head is characterized by the highest $\overline{H}$ factor and is constructed of only 24 modules.

\begin{figure}[ht]
    \centering
    \begin{subfigure}{}
        \centering
        \includegraphics[width=\linewidth]{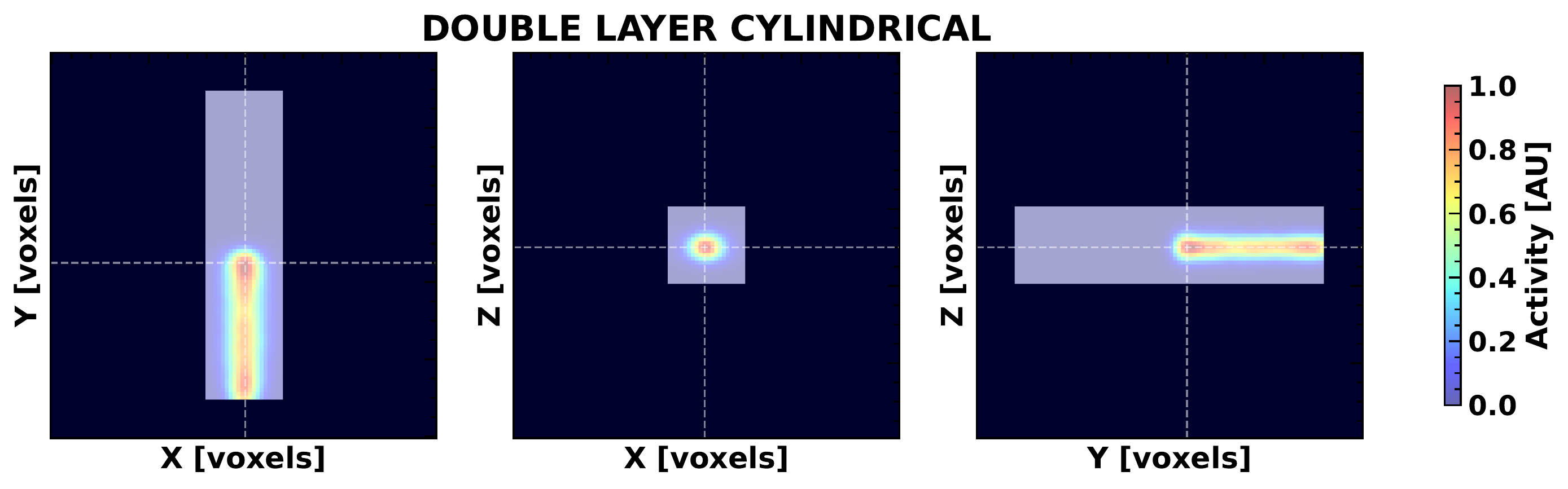}
    \end{subfigure}
    \hfill
    \begin{subfigure}{}
        \centering
        \includegraphics[width=\linewidth]{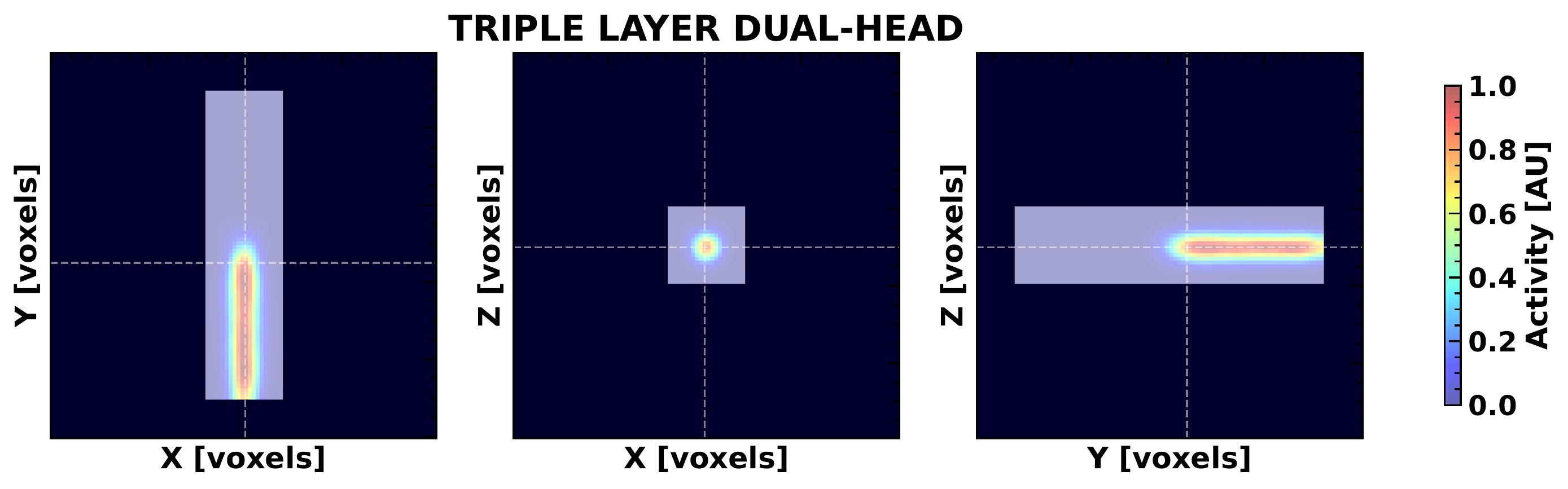}
    \end{subfigure}
    \caption{Example of the reconstructed PET images resulting from the SPB irradiation at 125.68 MeV and range of 100\,mm. The activity distributions reconstructed with 3 iterations are shown for the double layer cylindrical (top row) and triple layer dual-head (bottom row) geometries. Post-reconstruction 3D Gaussian smoothing of the activity distribution with $\sigma$ equal to 1 voxel was applied. The PMMA phantom size is 5$\times$20$\times$5~mm$^3$. The voxel size is 2.5$\times$2.5$\times$2.5~mm$^3$.}
    \label{fig:spb_100_recons_barrels}
\end{figure}

\begin{figure}
    \centering
    \begin{subfigure}{}
        \centering
        \includegraphics[width=\linewidth]{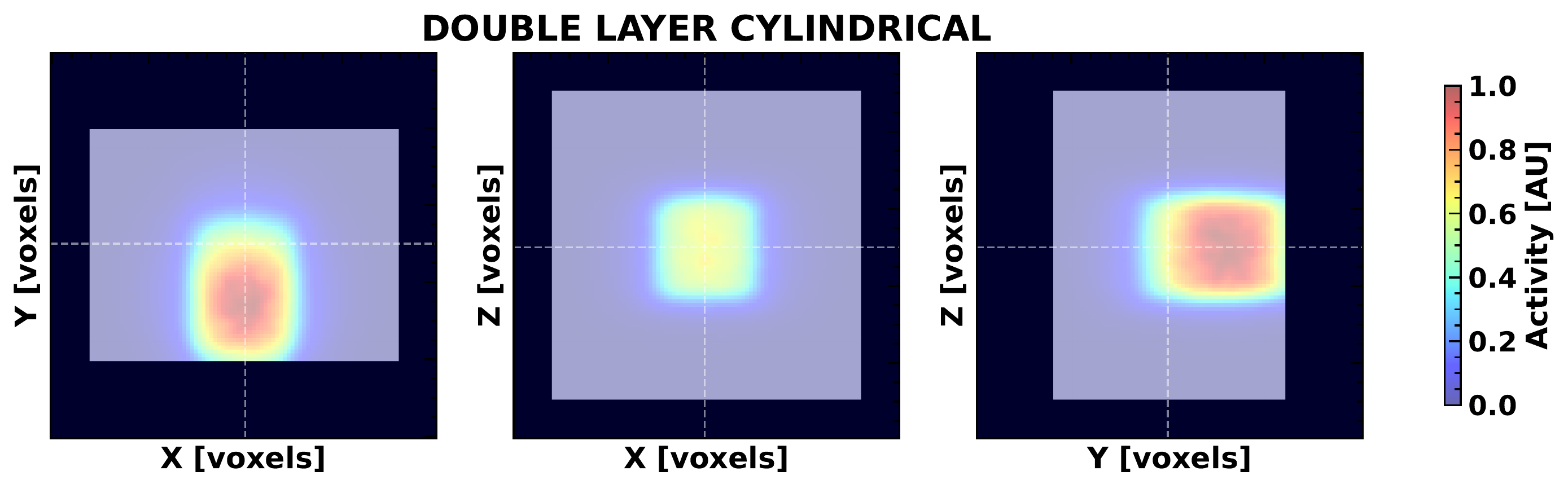}
    \end{subfigure}
    \hfill
    \begin{subfigure}{}
        \centering
        \includegraphics[width=\linewidth]{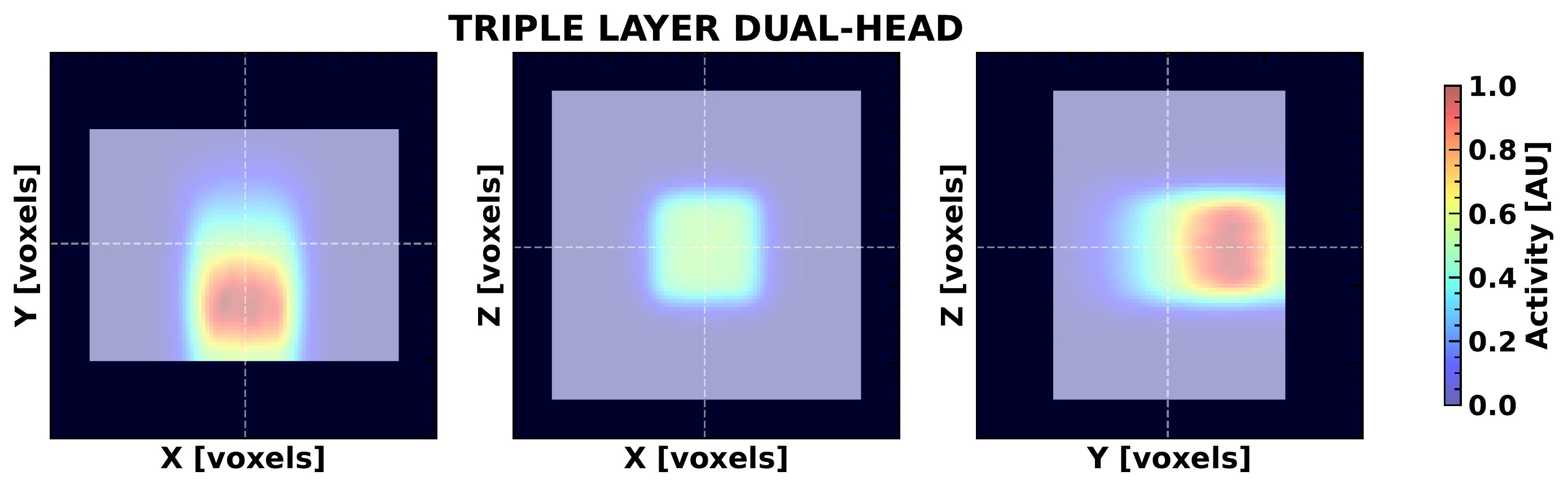}
    \end{subfigure}
    \caption{Example of the reconstructed PET images resulting from irradiation of the SOBP of 5$\times$5$\times$5\,cm$^3$ and 100\,mm range. The activity distributions reconstructed with 3 iterations are shown for the double layer cylindrical (top row) and triple layer dual-head (bottom row) geometries. Post-reconstruction 3D Gaussian smoothing of the activity distribution with $\sigma$ equal to 1 voxel was applied. The PMMA phantom size is 20$\times$15$\times$20~mm$^3$. The voxel size is 2.5$\times$2.5$\times$2.5~mm$^3$.}
    \label{fig:sobp_100_recons_dualheads}
\end{figure}

\begin{figure}[h!]
\begin{subfigure}
\centering
\includegraphics[width=0.5\linewidth]{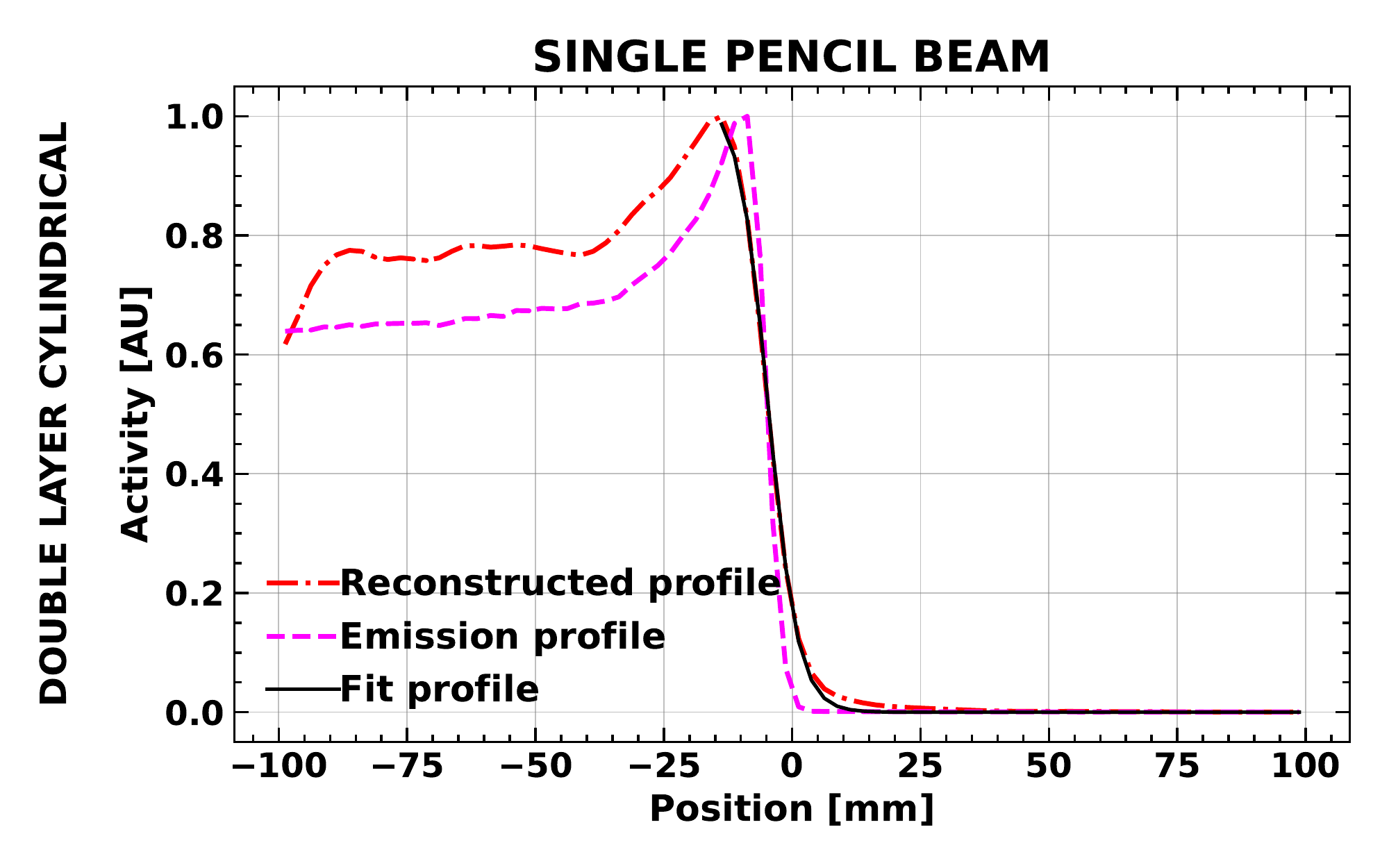}
\end{subfigure}
\hfill
\begin{subfigure}
\centering
\includegraphics[width=0.5\linewidth]{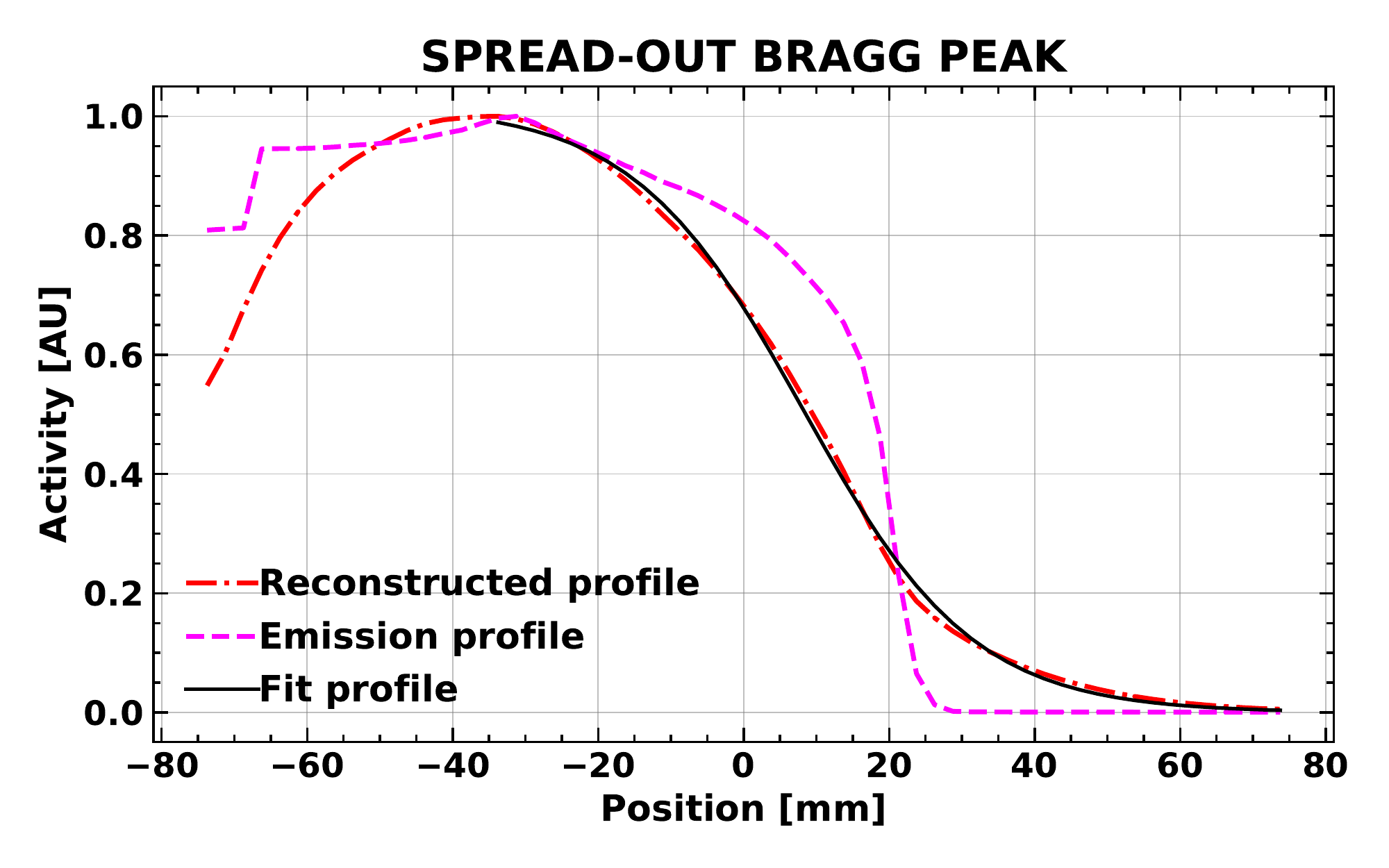}
\end{subfigure}
\hfill
\begin{subfigure}
\centering
\includegraphics[width=0.5\linewidth]{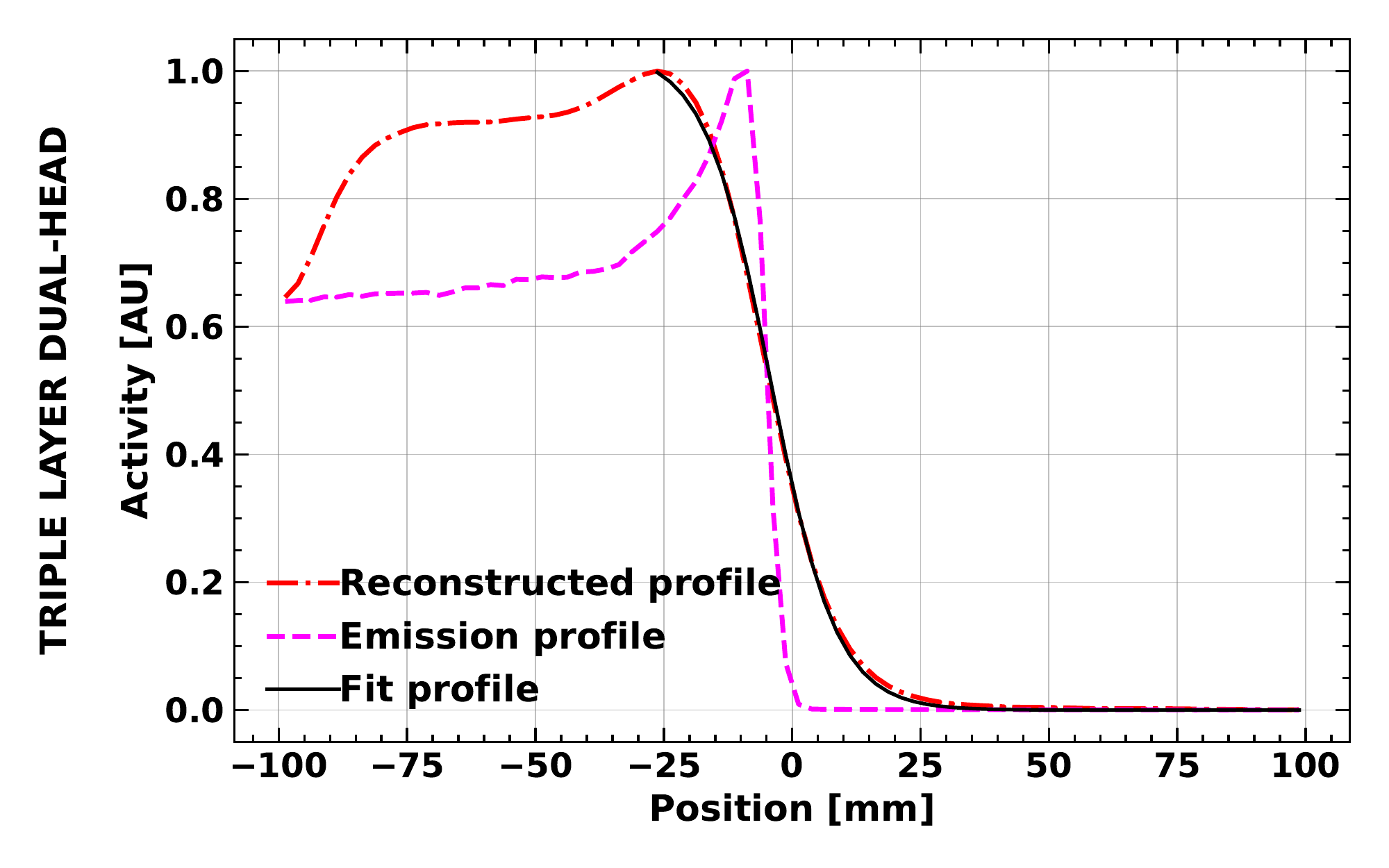}
\end{subfigure}
\hfill
\begin{subfigure}
\centering
\includegraphics[width=0.5\linewidth]{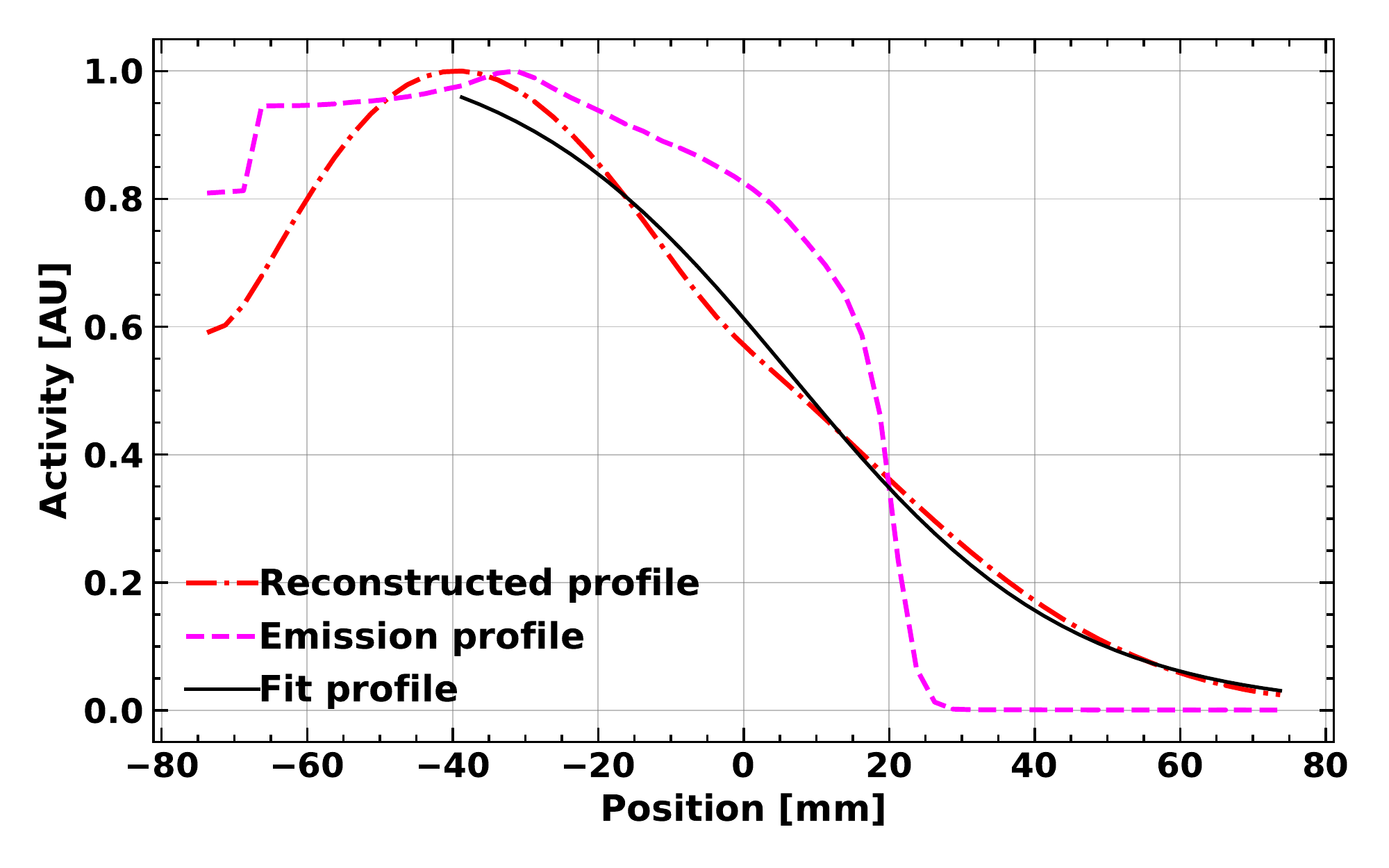}
\end{subfigure}
\caption[]{Examples of emission profiles, reconstructed activity profiles and sigmoid function fitted to the reconstructed profiles for SPB (left column) and SOBP irradiations (right column) for double layer cylindrical (top row) and triple layer dual-head (bottom row) geometry configurations.}
\label{fig:ranges}
\end{figure}

Reconstructed PET images from the double layer cylindrical and triple layer dual-head geometry, for SBP and SOBP irradiations, are shown in Fig.~\ref{fig:spb_100_recons_barrels} and Fig.~\ref{fig:sobp_100_recons_dualheads}, respectively. For both geometries, the relation between reconstructed images and the corresponding dose, production, and emission distributions, was previously shown in Borys et al. 2022~\cite{borys2022protheramon}. Furthermore, in Fig.~\ref{fig:ranges} we show an example of profiles taken through images reconstructed following the SPB and SOBP irradiations, together with the fitted sigmoid functions, as well as the corresponding emission profiles for comparison. As in Fig.~\ref{fig:spb_100_recons_barrels} and Fig.~\ref{fig:sobp_100_recons_dualheads}, only the double layer cylindrical and triple layer dual-head configurations are shown. 

Note that the fall-off reconstructed image profiles for the SBP irradiations are qualitatively more similar to the emission fall-offs than the fall-offs obtained from the SOBP irradiations. 

\subsection{Precision of range shift detection with various J-PET geometries}


\begin{figure}
\centering
\begin{subfigure}{}
  \centering
  \includegraphics[width=0.32\linewidth]{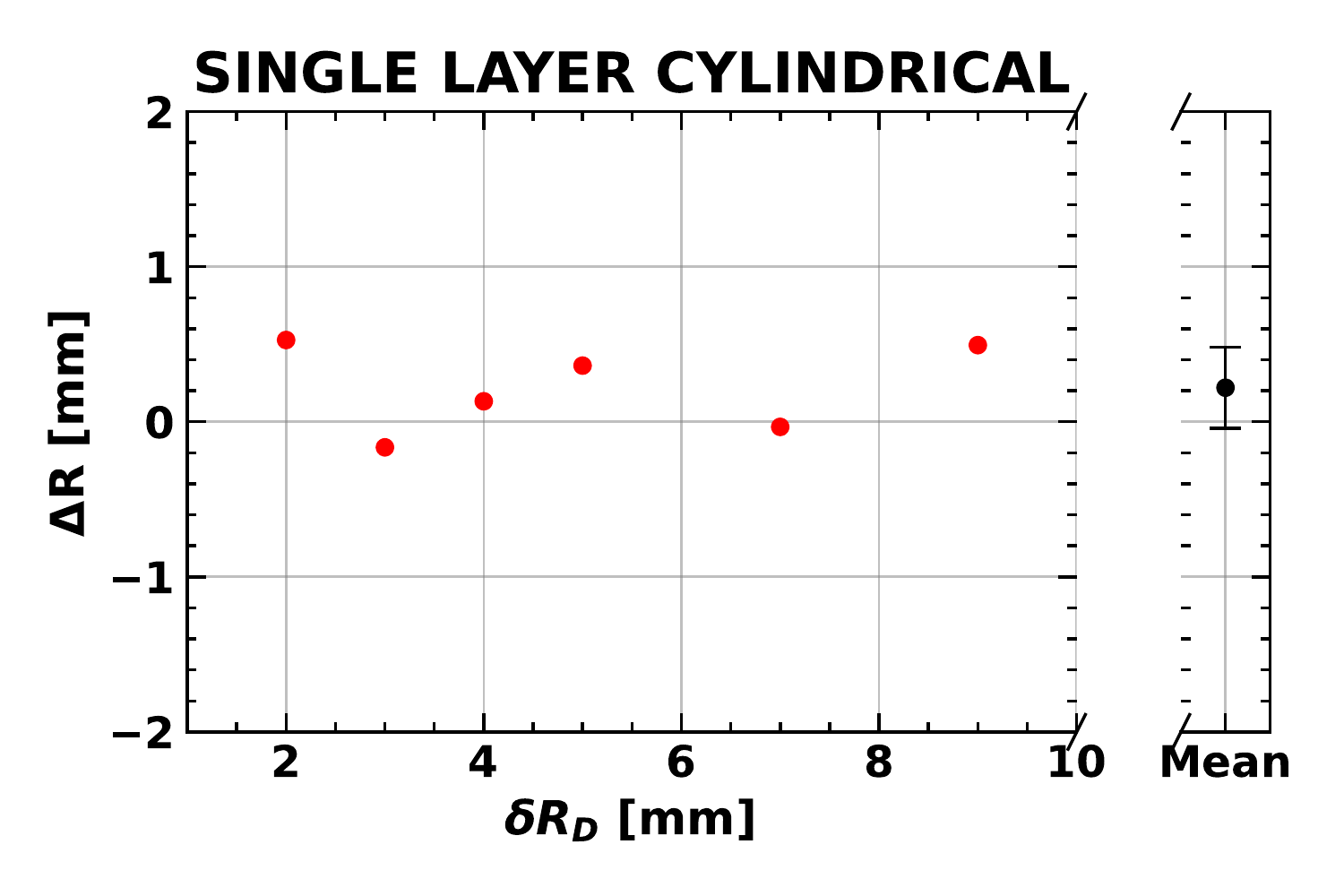}
  \includegraphics[width=0.32\linewidth]{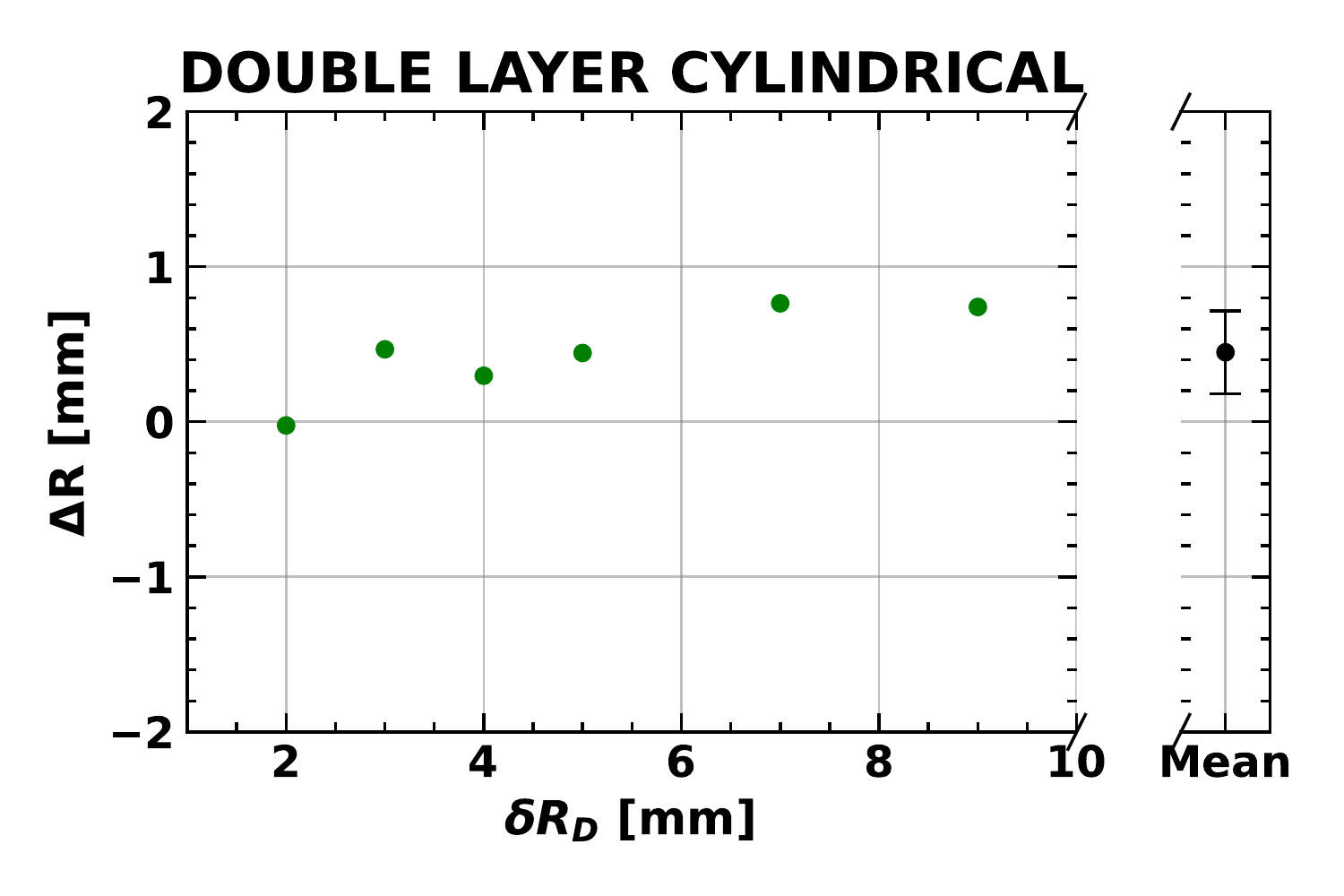}
  \includegraphics[width=0.32\linewidth]{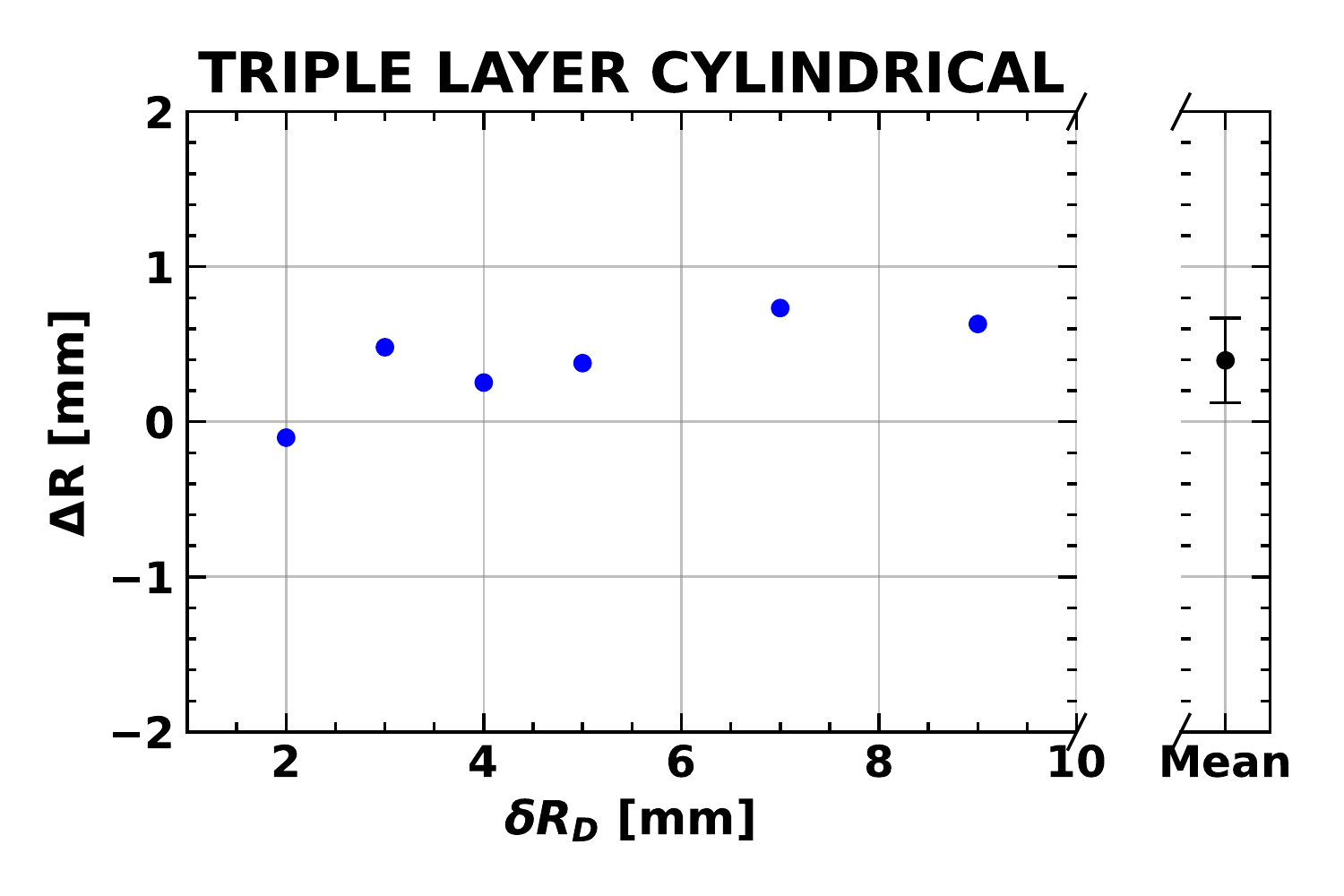}
\end{subfigure}
\begin{subfigure}{}
  \centering
  \includegraphics[width=0.32\linewidth]{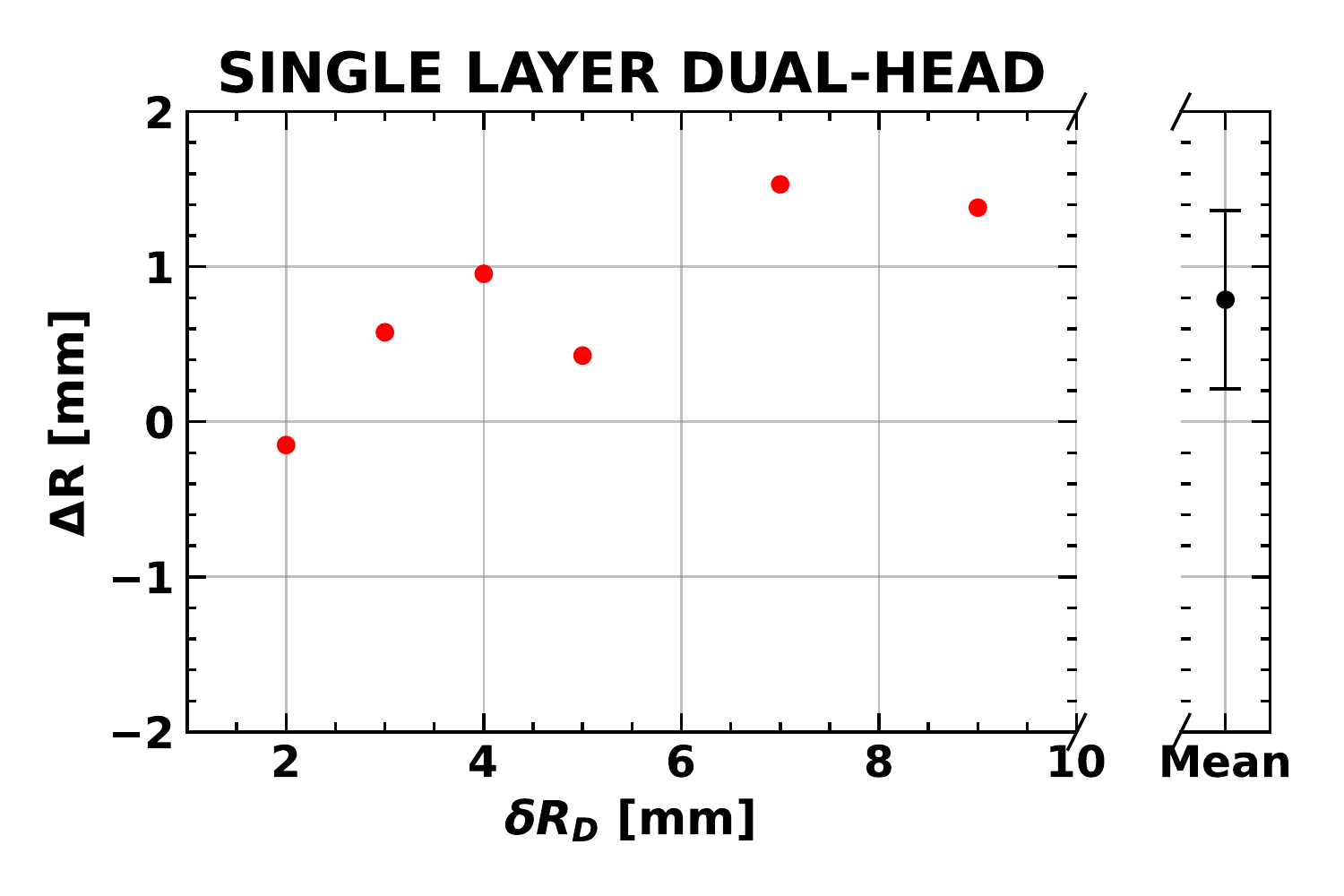}
  \includegraphics[width=0.32\linewidth]{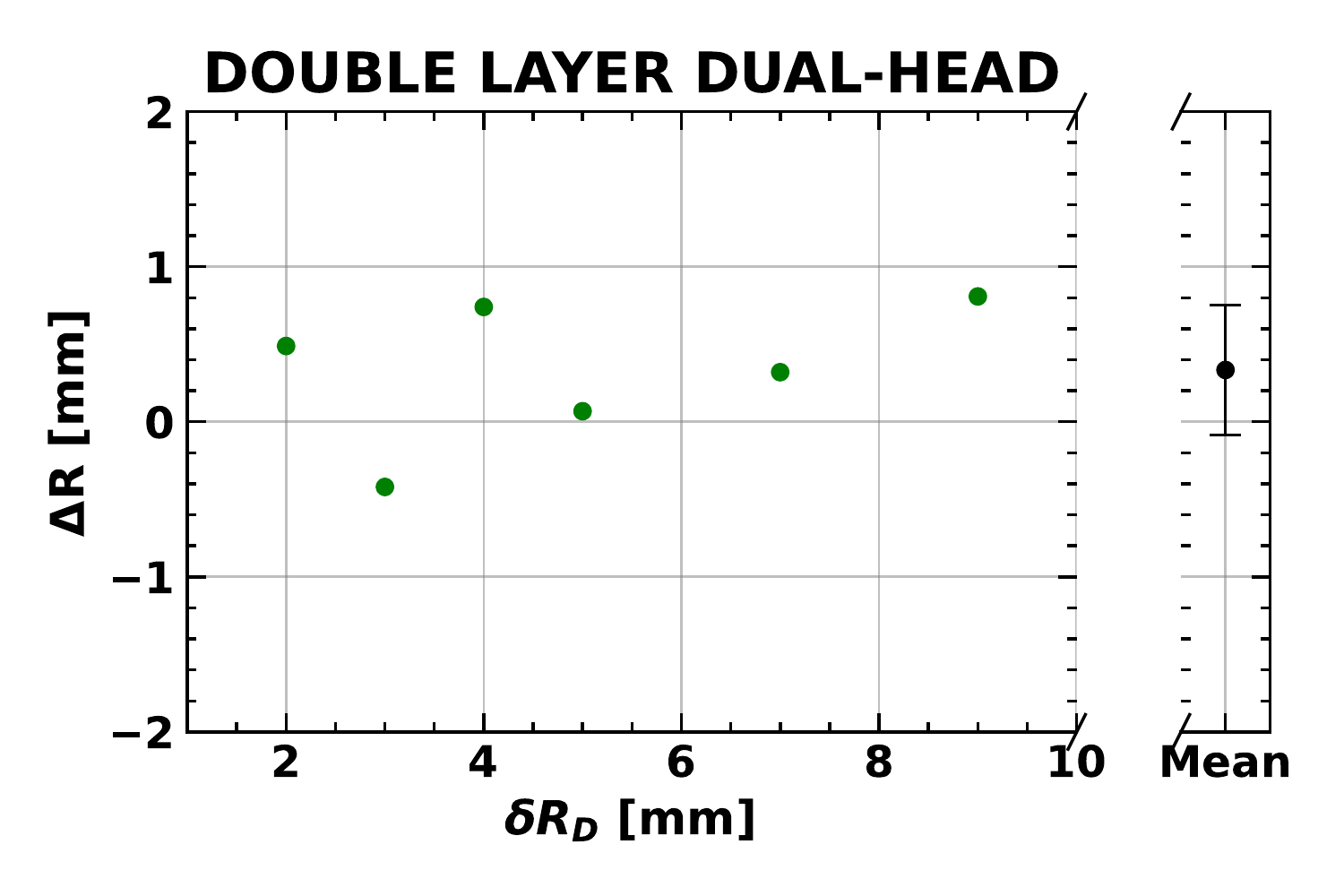}
  \includegraphics[width=0.32\linewidth]{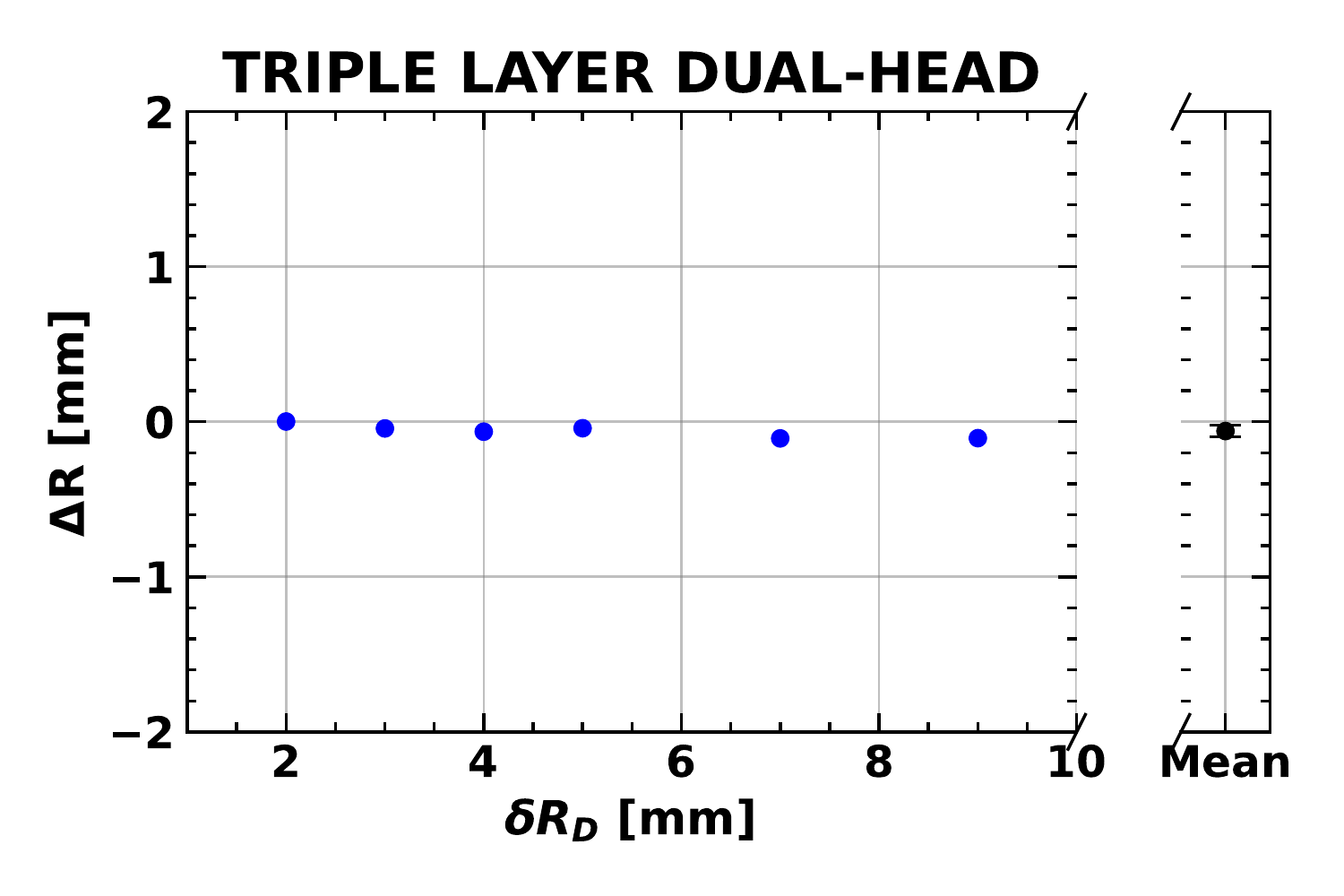}
\end{subfigure}
\caption{$\Delta R$ for all six geometries of the J-PET scanner investigated for SPB irradiations. Top row shows the results for cylindrical geometries and the bottom for the dual-head setups. The columns from left to right present the results for the single, double and triple layer scanner geometries.}
\label{fig:spb_means}
\end{figure}

\begin{figure}
\centering
\begin{subfigure}{}
  \centering
  \includegraphics[width=0.32\linewidth]{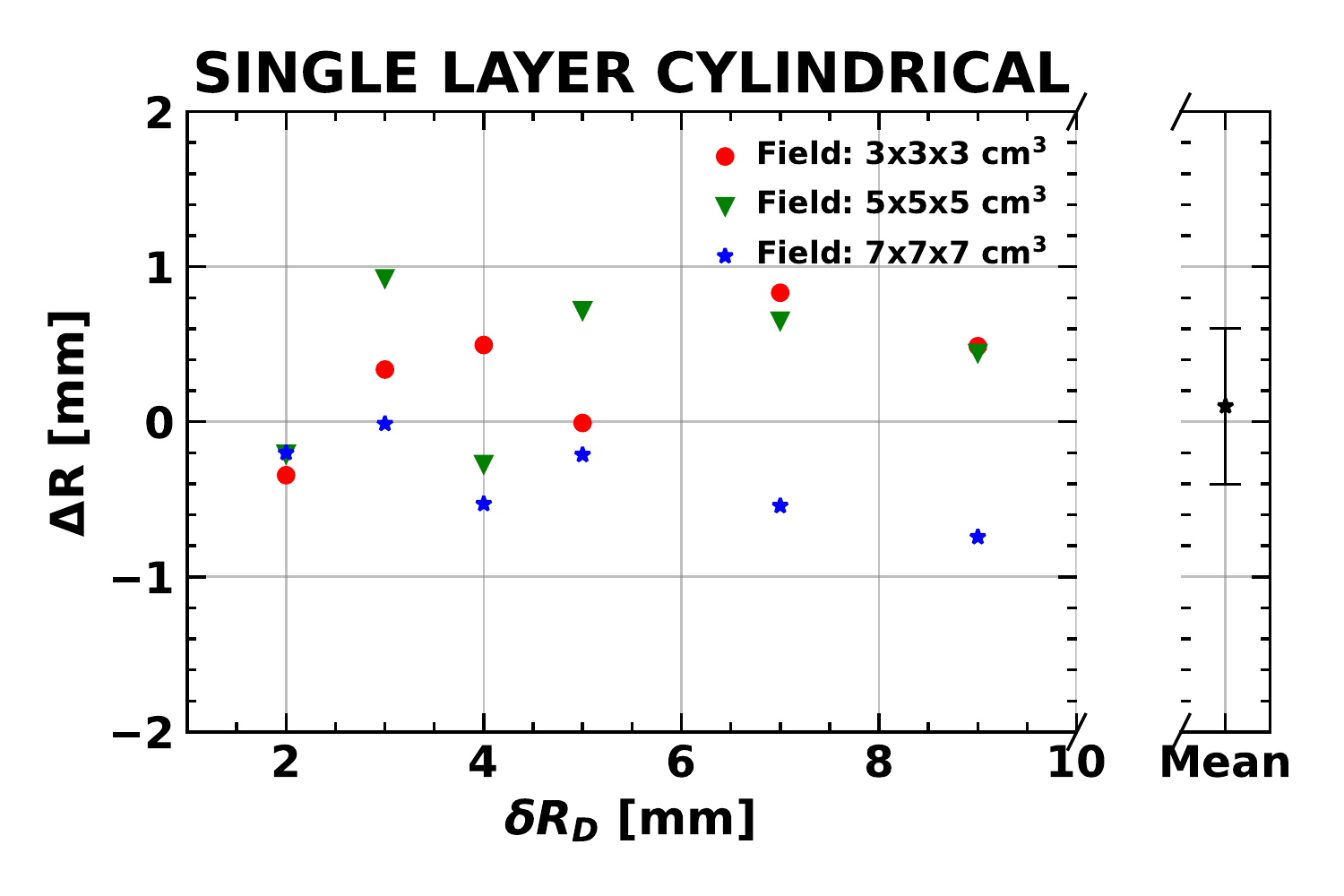}
  \includegraphics[width=0.32\linewidth]{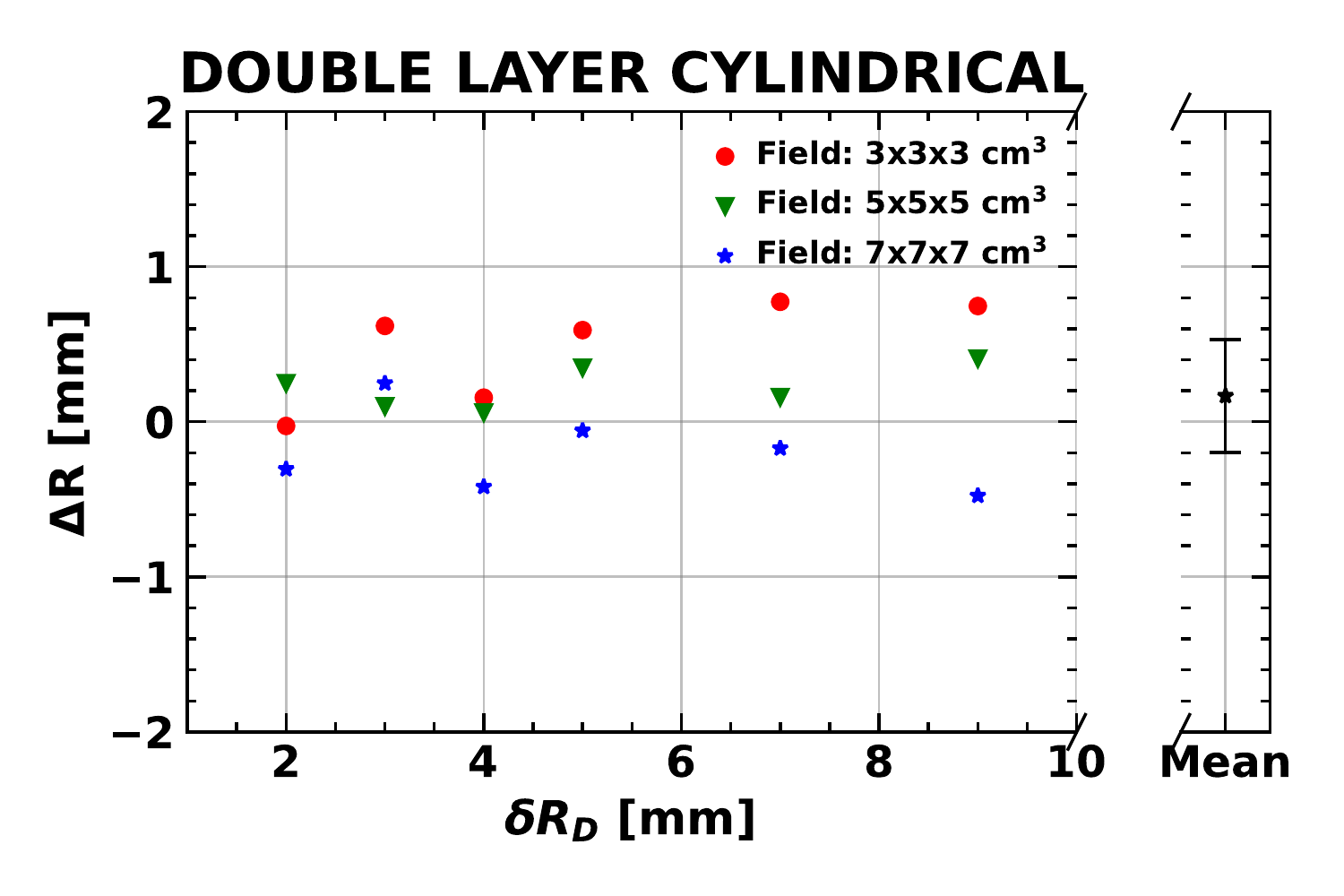}
  \includegraphics[width=0.32\linewidth]{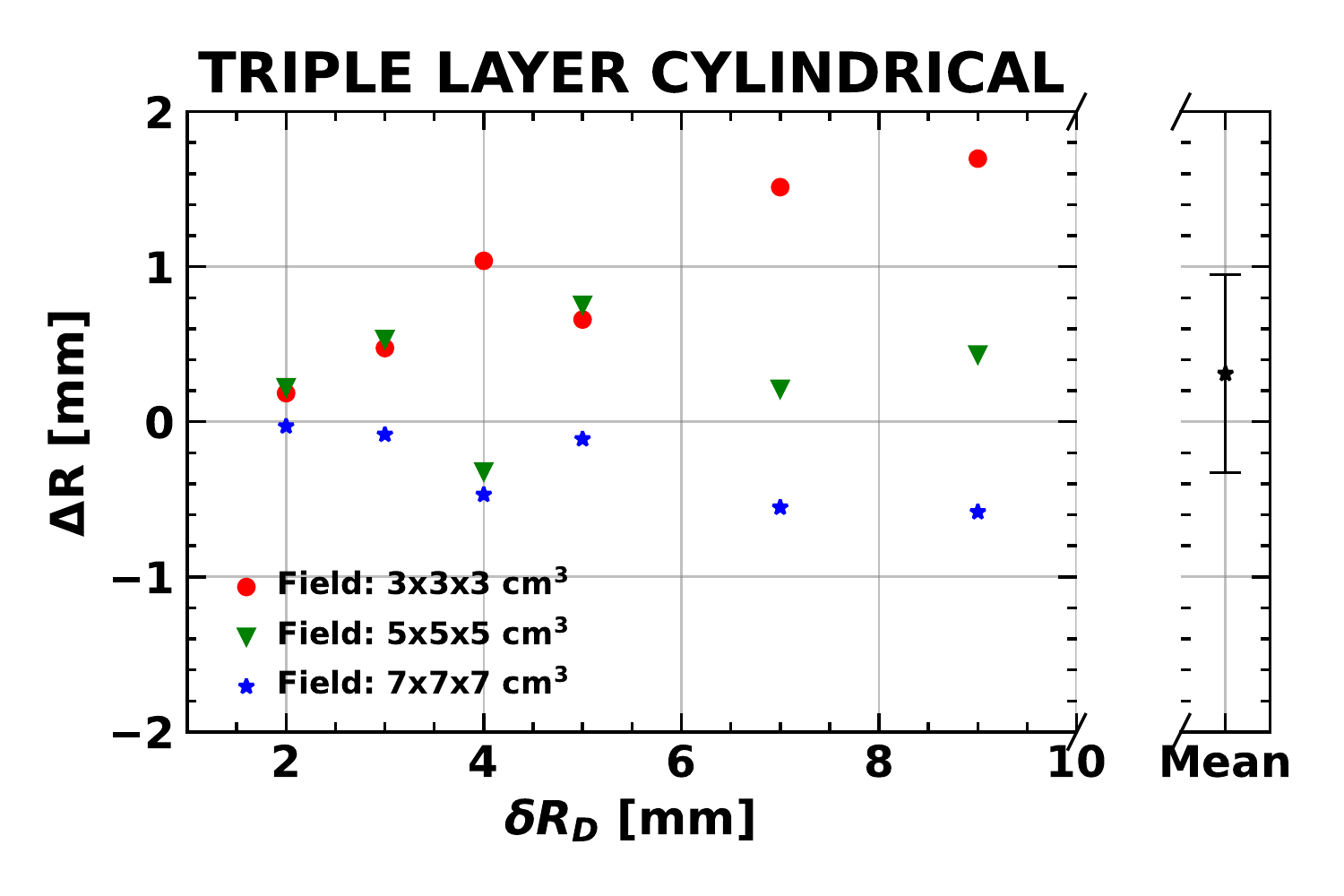}
\end{subfigure}
\begin{subfigure}{}
  \centering
  \includegraphics[width=0.32\linewidth]{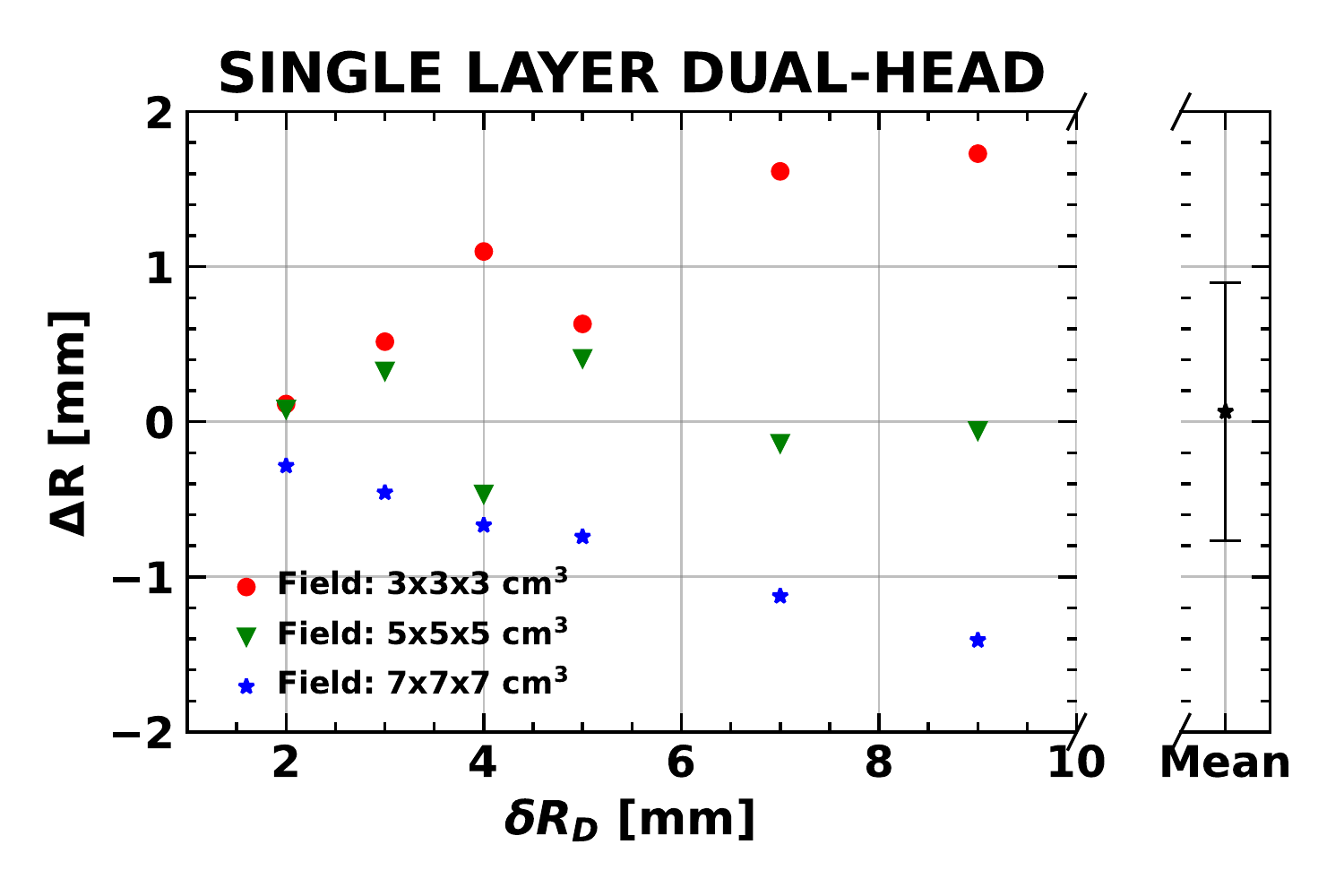}
  \includegraphics[width=0.32\linewidth]{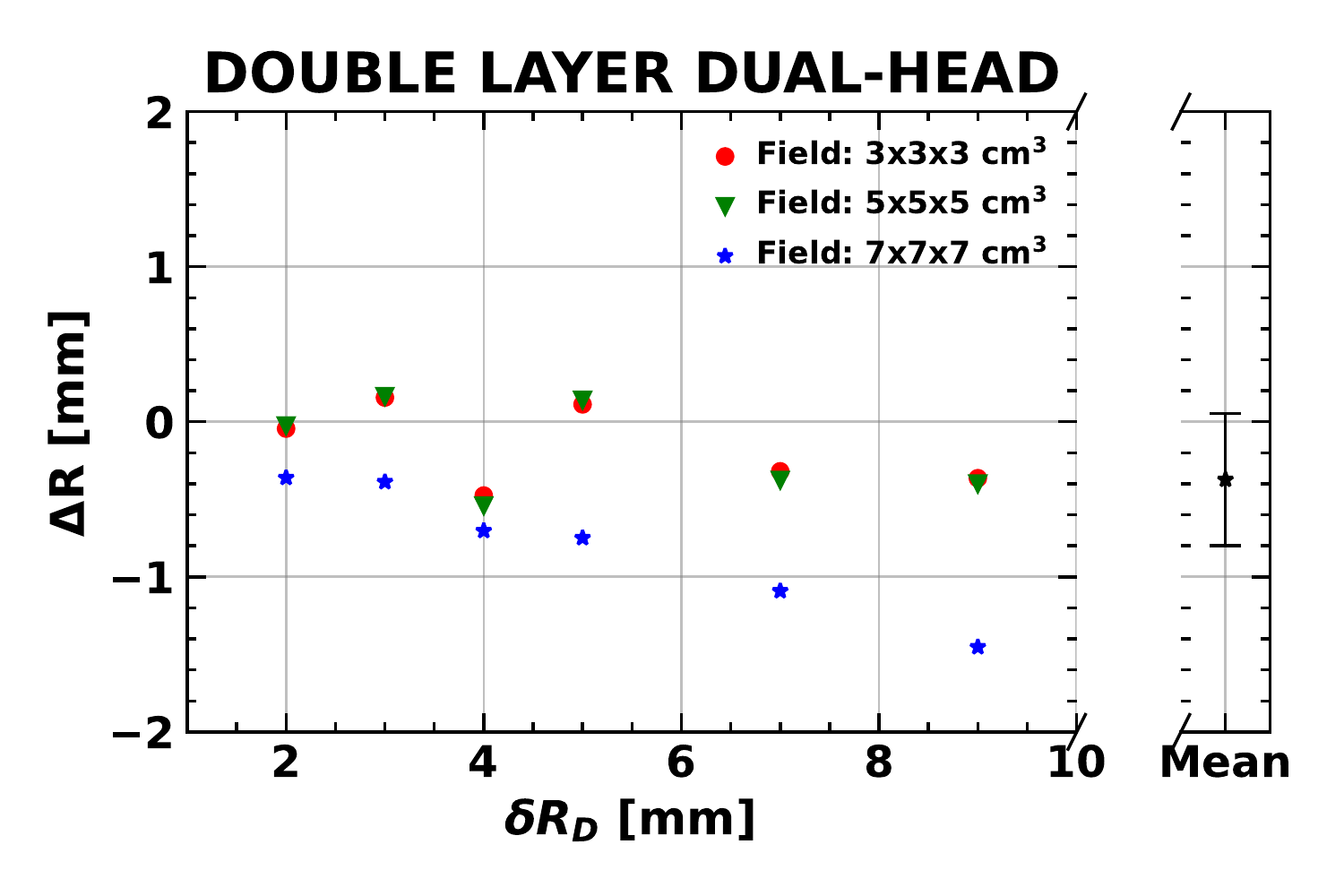}
  \includegraphics[width=0.32\linewidth]{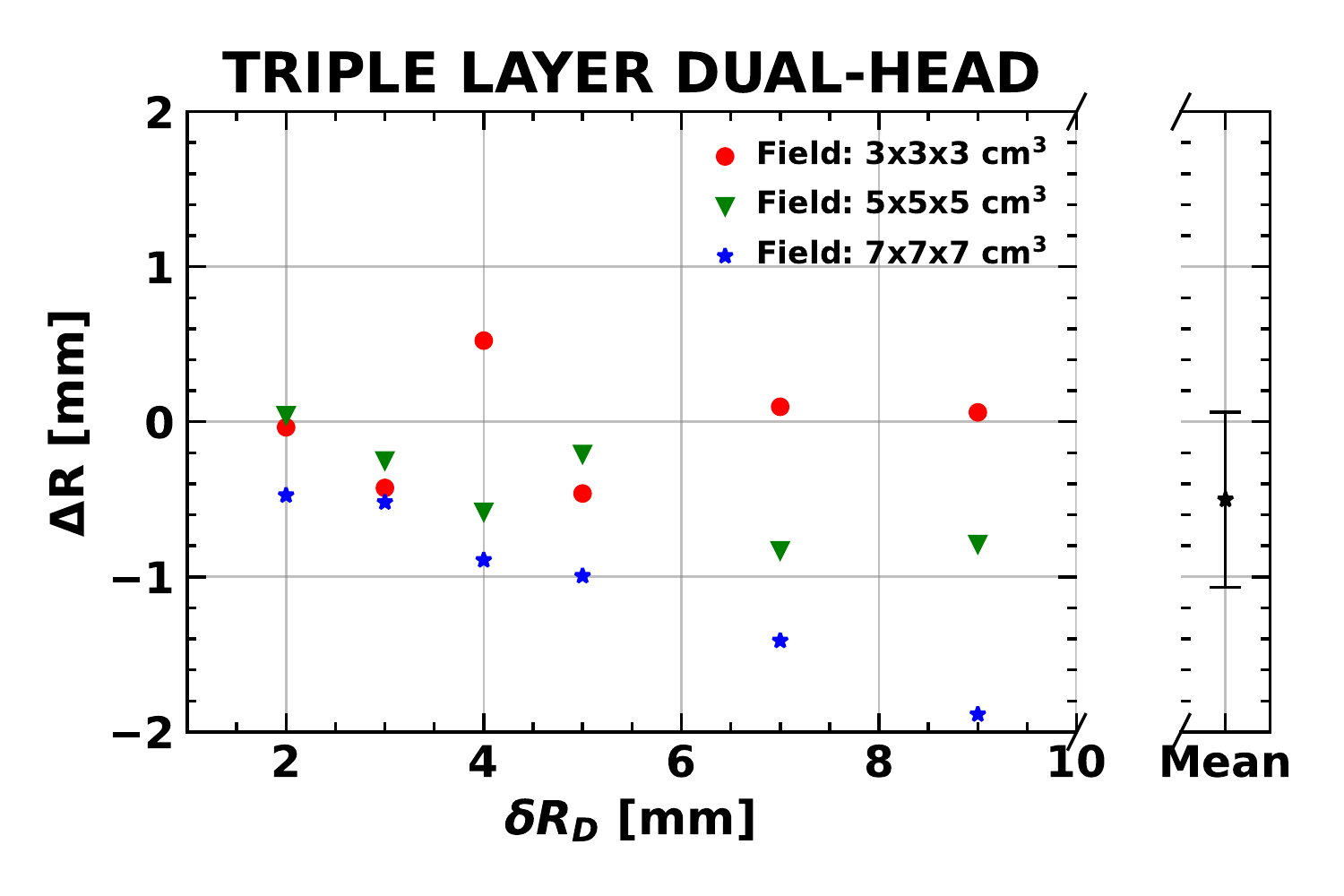}
\end{subfigure}
\caption{$\Delta R$ for all six geometries of the J-PET scanner investigated for SOBP irradiations with different fields size. Top row shows the results for cylindrical geometries and the bottom for the dual-head setups. The columns from left to right present the results for the single, double and triple layer scanner geometries.}
\label{fig:sobp_means}
\end{figure}

Fig.~\ref{fig:spb_means} and Fig.~\ref{fig:sobp_means} show $\Delta R$, the difference between the range shift in the dose and the range shift as measured from the reconstructed activity distributions after fitting the sigmoid function, for simulations of SPB and SOBP field irradiations, respectively.
The figures have six panels, each of them showing the results for one of the  investigated geometrical configurations of the J-PET scanner. For the SOBP irradiations, the results for three irradiation fields of different sizes are given. Additionally, in Table~\ref{tab:spb_sobp_means} calculated mean and standard deviation values are given separately for the SPB and SOBP irradiation. 

The mean of the differences in range shift, between the reference and the measurements made on reconstructed images, ranges from -0.37\,mm to 0.85\,mm. These values are smaller than the voxel size of the reconstructed PET image (isotropic 2.5\,mm). The standard deviation of $\Delta R$, which we associate with the precision of the range detection, is below 1\,mm for both, SPB and SOBP irradiations. For all the investigated geometries, higher precision is observed for the SPB than SOBP. The best precision for the SPB is found for the triple layer dual-head and the worst for the single layer dual-head. All cylindrical configurations show similar precision for the SPB investigation. Among the dual-head setups, the single layer geometry has the worst results for both SOBP and SPB studies. 

Comparing the geometries with the same number of modules, for the SPB the best range detection precision was found for the triple layer dual-head, whereas for the SOBP the double layer dual-head configurations was found to be superior.

\begin{table}[!ht]
\caption{Calculated distances between measured and reference difference (SOBP study).}
\centering
\begin{tabular}{||c c c c c||} 
\hline
 & \multicolumn{2}{c}{SPB study}  & \multicolumn{2}{c||}{SOBP study}\\
\cline{2-5}
Setup & $\overline{\Delta R}$ [mm]
& $\sigma_{\Delta R}$  [mm] & $\overline{\Delta R}$ [mm]
& $\sigma_{\Delta R}$ [mm] \\
 \hline \hline
Single layer cylindrical & 0.22 & 0.26 & 0.10 & 0.50\\
 \hline
Double layer cylindrical & 0.45 & 0.27 & 0.17 & 0.36\\
 \hline
Triple layer cylindrical & 0.40 & 0.27 & 0.31 & 0.64\\
 \hline
Single layer dual-head & 0.79 & 0.58 & 0.07 & 0.83\\
 \hline
Double layer dual-head & 0.33 & 0.42 & -0.37 & 0.43\\
 \hline
Triple layer dual-head & -0.06 & 0.04 & -0.05 & 0.56 \\ 
 \hline
\end{tabular}
\label{tab:spb_sobp_means}
\end{table}

\section{Discussion}


We proposed and compared six geometries of the modular J-PET scanner (three dual-head and three cylindrical) in terms of sensitivity and precision for proton beam range measurements. We used the \protheramon software framework to conduct Monte Carlo simulations of SPB and SOBP irradiations in a uniform PMMA phantom. We reconstructed the resulting \betaplus activity distributions using the CASTOR software. 
The number of detected coincidences per primary proton ($\eta$) varies between the investigated J-PET scanners from $0.4\cdot10^{-5}$ to $4.7\cdot10^{-5}$ for a SPB and from $0.2\cdot10^{-5}$ to $2.0\cdot10^{-5}$ for a SOBP. The sensitivity of the scanners relative to single-layer cylindrical geometry, $H_{geom}$, for the SPB and SOBP irradiation plans approaches 5 for triple-layer cylindrical geometry. Quantitative analysis was conducted to assess the precision of range detection with different scanner geometries, which for the investigated SPB and SOBP irradiations and using uniform PMMA phantoms was found to be below 1\,mm. 

The $\Delta R$ values, the difference between the range shift in the dose and the range shift as estimated from the reconstructed activity distributions, are mostly smaller than the expected J-PET resolution, which is at the level of a few millimeter~\cite{moskal2021simulating}. Therefore, it can be concluded that following experimental verification, in principle, all investigated configurations could be considered for practical application in proton range monitoring. However, it should be stressed, that the investigations presented here were performed in a uniform phantom. Further studies that are currently under review~\cite{Brzezinski2023_PMBsubmitted} are required to assess the feasibility of range detection for heterogeneous Intensity-Modulated Proton Therapy treatment plans in non-uniform patient tissue. Therefore, small differences between the investigated geometries observed here for simplified quality assurance settings may result in much more significant differences in a clinical setting. In this context, both precision and sensitivity should be considered essential factors for geometry optimization, taking into account that the general rule for PET imaging is that greater statistics will improve the reconstructed image quality.

The presented results, considering both sensitivity and precision, as well as the cost-effectiveness of J-PET based configurations, indicate that the double-layer cylindrical and triple-layer dual-head configurations are the most promising for in-room/off-beam and inter-spill/in-beam applications, respectively. The triple layer dual-head geometry has the greatest efficiency factor $\overline{H}$ (as it is shown in Table \ref{tab:efficiences}) among the scanners with 24 modules. This is the number of modules currently available and undergoing commissioning with the modular prototype of the system. The double-layer cylindrical geometry is the envisioned final Total-Body J-PET geometry~\cite{moskal2021simulating} and would potentially benefit from the experience of the J-PET group in scanner construction and future operation. Note that the double-layer setup has a sensitivity increase of 300\% with respect to the single-layer setup, while the triple-layer setup has a sensitivity increase of about 70\% with respect to the double-layer configuration, pointing to the double-layer scanner as the most cost-effective configuration.

Direct comparison with other PET range monitoring systems in not straightforward due to the differences in the experimental setup, e.g. phantom size or irradiation plans. Recently, the mobile PET system DoPET, developed at the University of Pisa, Italy \cite{kraan2019analysis,topi2019monitoring}, has been investigated for the application of range monitoring in proton therapy. Various phantoms were irradiated and PET signal was acquired immediately after the irradiation for five minutes, mimicking the in-room range monitoring approach. Their experiments and Monte Carlo simulations with FLUKA~\cite{battistoni2015overview,augusto2018overview} revealed that the efficiency factor (number of coincidences per primary protons) is at the level of $\eta$=2.85\cdot10$^{-5}$. The double- and triple-layer J-PET scanners investigated in this study have the sensitivity of the same order of magnitude as the DoPET system. However, a comparison of the precision in range measurement is challenging, considering the different irradiation and PET acquisition scenarios (phantoms, treatment plan, acquisition protocol), and is beyond the scope of this manuscript. In comparison to DoPET, the J-PET systems show similar sensitivity, while its advantages are the cost-effectiveness and modular design, which enable construction systems capable of various PET acquisition scenarios and facilitate installation in treatment rooms of different designs.

The uncertainties of the presented simulations study are related to the assumptions made. 
We applied an artificial discretization of the plastic strip into a hundred 5-mm long pseudo-crystals, which in the first approximation is in agreement with the expected resolution of the currently produced 3rd generation J-PET scanner that uses Wave Length Shifters (WLS) that offer improved time resolution~\cite{smyrski2017measurement,moskal2021simulating}. Minor uncertainties may relate to the physics modelling used for the simulation of the activity production and propagation of the 511 keV annihilation photons~\cite{borys2022protheramon}. Based on the clinical protocols used in CCB Krakow proton therapy centre, we have also assumed 2 minutes PET acquisition time~\cite{Brzezinski2023_PMBsubmitted}, while the detected PET signal may substantially vary when modifying acquisition time. The uncertainties related to the PET image reconstruction are related to e.g. sensitivity correction, normalization, attenuation correction, number of iterations used in the reconstruction, post-reconstruction image filtering. 
To simplify the presented preliminary analysis, we have omitted the propagation of the uncertainty related to the fitting of the activity profile fall-off to the $\delta R$ and $\sigma R_D$ values. 
The simulated phantom irradiations were performed with relatively high doses of 4~Gy (SBP) and 8 Gy (SOBP), characteristic of the hypofractioned treatments, where range monitoring is of particular importance. Range monitoring of patients with 2~Gy fraction doses will result in lower coincidence statistics and is being further investigated in~\cite{Brzezinski2023_PMBsubmitted}.

For the assessment and characterization of diagnostic PET scanners, NEMA norms \cite{NEMA:2018} are used. We propose that the sensitivity and precision analysis presented here be the first step towards introducing similar norms for the evaluation of PET scanners for proton beam range monitoring. This evaluation should, furthermore, consider such aspects as cost-effectiveness and the suitability of the technology for intra-treatment PET imaging. We believe that the \protheramon framework, offering a standardized simulation and image reconstruction environment, may be helpful for reliable comparison of different setups. A consensus and guidelines for the evaluation of PET-based range monitoring techniques would be of great benefit in fostering future developments in proton beam range monitoring and in its translation into the clinic.


\section{Conclusions}
\label{sec:conclusions}

In this paper, the feasibility of the J-PET detector for PET-based proton beam therapy range monitoring was investigated. Six different scanner geometries were tested by means of Monte Carlo simulations. Relative efficiency and range assessment precision were computed in order to find the optimal geometrical configuration.

The study reveals that considering the sensitivity, precision and cost-effectiveness of different approaches, the most promising for the clinical application are the double-layer cylindrical and triple-layer dual-head configurations, dedicated for in-room/off-beam and inter-spill/in-beam applications, respectively. Among the scanners with 24 modules of the J-PET detector, the best results were obtained with the triple-layer dual-head system. All the systems show the feasibility of range assessment with precision at the level of 1 mm for both SPB and SOBP irradiations. Experimental validation of the presented results is needed and ongoing.

\section*{Acknowledgments}
\label{sec:acknowledgments}

This work was supported by the National Centre for Research and Development (NCBiR), grant no. LIDER/26/0157/L-8/16/NCBR/2017. 
Calculations were performed on the Ziemowit computer cluster in the Laboratory of Bioinformatics and Computational Biology at the Biotechnology Centre, the Silesian University of Technology, created in the EU Innovative Economy Programme POIG.02.01.00–00-166/08 and expanded in the POIG.02.03.01–00-040/13 project. 
Development of \protheramon was supported by the National Centre for Research and Development (NCBiR), grant no. LIDER/43/0222/L-12/20/NCBR/2021.
We acknowledge also support by the Foundation for Polish Science through the TEAM POIR.04.04.00-00-4204/17 program, the National Science Centre of Poland through grant 2021/42/A/ST2/00423, and the SciMat and qLife Priority Research Areas budget under the program {\it Excellence Initative - Research University} at the Jagiellonian University, and Jagiellonian University project no. CRP/0641.221.2020.

\section*{References}
\label{sec:references}

\bibliography{references_nano_add}

\end{document}